\documentclass[%
 reprint,
 amsmath,amssymb,
 aps,
]{revtex4-2}

\usepackage{graphicx}
\usepackage{dcolumn}
\usepackage{bm}
\usepackage{amsmath}  
\usepackage{physics}  
\usepackage{slashed}  
\usepackage{amssymb}

\usepackage{xcolor}

\newcommand{\zh}{\hat{z}}
\newcommand{\p}{\partial}
\newcommand{\del}{\delta}
\newcommand{\e}{\eta}
\newcommand{\ep}{\epsilon}
\newcommand{\zb}{\overline{\zeta}}
\newcommand{\z}{\zeta}

\newcommand{\s}{\sigma}
\newcommand{\m}{\mu}
\newcommand{\n}{\nu}
\newcommand{\rh}{\rho}
\newcommand{\al}{\alpha}
\newcommand{\mn}{\mu\nu}

\newcommand{\g}{\gamma}
\newcommand{\om}{\omega}

\newcommand{\gd}{\dot{\g}}

\newcommand{\egd}{\e \cdot \dot{\g}}

\newcommand{\kgd}{k \cdot \dot{\g}}
\newcommand{\qgd}{q \cdot \dot{\g}}
\newcommand{\kg}{k \cdot \g}
\newcommand{\kx}{k \cdot x}
\newcommand{\eq}{\e \cdot q}
\newcommand{\lgd}{l \cdot \dot{\g}}

\newcommand{\gdd}{\Ddot{\g}}

\begin{document}

\preprint{APS/123-QED}

\title{First look at continuous spin gravity: Time delay signatures}

\author{Shayarneel Kundu}
\email{skundu4@stanford.edu}
\affiliation{
 SLAC National Accelerator Laboratory, 2575 Sand Hill Road, Menlo Park, CA 94025, USA
}
\affiliation{Department of Physics, Stanford University, Stanford, CA, 94305, USA.}
\author{Philip Schuster}%
 \email{schuster@slac.stanford.edu}
\affiliation{
 SLAC National Accelerator Laboratory, 2575 Sand Hill Road, Menlo Park, CA 94025, USA
}
 \author{Natalia Toro}%
 \email{ntoro@slac.stanford.edu}
\affiliation{
 SLAC National Accelerator Laboratory, 2575 Sand Hill Road, Menlo Park, CA 94025, USA
}

\date{\today}

\begin{abstract}
We consider the possibility that gravity is mediated by "continuous spin" particles, i.e.~ massless particles whose invariant spin scale $\rho_g$ is non-zero. In this case, the primary helicity-2 modes of gravitational radiation on a Minkowski background mix with a tower of integer-helicity partner modes under boosts, with $\rho_g$ controlling the degree of mixing. We develop a formalism for coupling spinless matter to continuous spin gravity at linearized level. Using this formalism, we calculate the time-delay signatures induced by gravitational waves in an idealized laser interferometer detector.  The fractional deviation from general relativity predictions is  $O(\rho_g/\omega)$ for gravitational wave frequencies $\omega >\rho_g$, and the effects of waves with $\omega \lesssim \rho_g$ are damped. The precision and low frequency ranges of gravitational wave detectors suggest potential sensitivity to spin scales at or below $\sim 10^{-14}$ eV at ground-based laser interferometers and $\sim 10^{-24}$ eV at pulsar timing arrays, motivating further analysis of observable signatures.
\end{abstract}

\maketitle


\section{Introduction}\label{1}

Lorentz symmetry and quantum mechanics impose powerful constraints on interacting massless particles in flat space. In particular, considering particles with Lorentz-invariant helicity $h$, the classic theorems of Weinberg and Witten \cite{Weinberg:1964ev, Weinberg:1964ew, Weinberg:1965rz, Weinberg:1980kq} show that only $|h|\leq 2$ can mediate long-range forces, with the perturbative graviton and gravitino as the only consistent possibilities with $|h|\ge 1$, and helicities $\pm1$ necessarily coupling to a conserved current. In this way, the mathematical consistency of quantum theories expanded about flat space appears to be an effective guiding principle to constructing theories of nature. 

There is, however, an important gap in the above arguments: they assume that massless particles carry Lorentz-invariant helicity, which is \emph{not} required by Lorentz-covariance of the theory. Theories where this constraint is lifted are known as ``continuous spin particle'' or CSP theories, and are characterized by a ``spin-scale'' $\rho$ with units of momentum \footnote{Mathematically, this class of state corresponds to a non-vanishing spin Casimir  $W^2=-\rho^2$, where $W^\mu$ is the Pauli-Lubanski pseudo-vector. This possibility was first identified in Wigner's pioneering classification of particles' spins in the 1930's \cite{Wigner:1939cj}.}. Though the theory is not yet fully developed, recent work suggests that CSPs may \emph{also} be able to mediate long-range interactions --- and that, at energies much larger than $\rho$, they become indistinguishable from familiar theories whose massless particle excitations have Lorentz-invariant helicity (e.g. perturbative GR about flat space and electromagnetism).  In particular, two of the authors generalized the Weinberg soft theorems to non-zero spin-scale in \cite{Schuster:2013pxj, Schuster:2013vpr}, finding that in all cases the helicity modes $h> 2$ decouple at mode energies (frequencies) $\gg \rho$. They also found in \cite{Schuster:2014hca} a gauge field theory for free CSPs that recovers familiar helicity actions in the $\rho\to0$ limit, and with Zhou \cite{Schuster:2023xqa} described interactions of these $\rho\neq 0$ gauge fields with a matter particle, for the cases where high-energy limits are dominated by either scalar-like interactions of the $h=0$ mode or gauge-boson-like interactions of the $h=\pm 1$ modes.  We also note with interest recent progress formulating CSP interactions with spinor-helicity techniques in \cite{Bellazzini:2024dco}, which could offer a complementary path to building up interacting theories with non-zero $\rho$.    Taken together, these results suggest that the known gauge theories of the Standard Model and General Relativity could just be approximate limits of a more complete class of theories with small but non-zero spin scale.

In this paper, we extend the treatment of \cite{Schuster:2023xqa} to describe the linearized coupling to matter of gravitons with $\rho \neq 0$, and use it to compute the response of an idealized gravitational wave detector. This application is of considerable interest for two reasons: First, gravitational wave detection explores the low-frequency domain, and does so with precision. Because theories with non-zero $\rho$ behave just like their $\rho \to 0$ counterparts at high frequencies $\omega \gg \rho$ in natural units, but introduce novel effects at frequencies of order $\rho$, kHz-frequency gravitational wave signals are deformed by $O(1)$ for $\rho$ as small as $10^{-14}$ eV.  Second, this problem is well suited to the limitations of presently developed formalism, which is most reliable in predicting the interactions with matter of ``on-shell'' CSPs---i.e.~plane wave solutions to the vacuum equations of motion---and is limited to leading order in Newton's gravitational constant, $G_N$.  Both of these are excellent approximations in the case of gravitational wave detection. Possible modifications from non-zero $\rho$ to gravitational wave \emph{production} are also interesting --- although here the leading order approximation in $G_N$ is well known to be inadequate, the results of \cite{Schuster:2023xqa,Schuster:2013vpr} imply parametric suppression of source effects relative to detection effects, as discussed in Section \ref{strainrho}, justifying our focus here on the well-controlled effect of non-zero $\rho$ for gravitational wave detection. 

Gravitational wave detection in a laser interferometer relies on \emph{anti-correlated, time-varying phase shifts} in laser light reflected in the two arms to measure the strain of the gravitational wave. Because the formalism of \cite{Schuster:2023xqa} treats matter as a particle rather than a field, it is useful to think of the phase-shift in each arm as a \emph{time-delay} (or advance) in the return of an individual photon due to the gravitational wave. 
In section \ref{2}, we motivate this time-delay observable, define it precisely, and discuss its computation in a worldline formalism of linearized General Relativity. In section \ref{3}, we  summarize the CSP field theory formalism needed to generalize this approach to $\rho_g \neq 0$, and we compute the $\rho_g -$dependent corrections to the gravitational ``time-delay."  In section \ref{4}, we outline related computations needed to make direct contact with the study of gravitational radiation in ongoing and future experiments. We note that a range of other authors has considered various aspects of CSP physics over the last decade, including descriptions of free theories \cite{Buchbinder:2020nxn, Burdik:2019tzg, Buchbinder:2019esz, Alkalaev:2018bqe, Rivelles:2018tpt, Buchbinder:2018yoo, Alkalaev:2017hvj, Khabarov:2017lth, Bekaert:2017khg, Najafizadeh:2017tin, Zinoviev:2017rnj, Metsaev:2018lth, Rivelles:2016rwo, Bekaert:2015qkt, Rivelles:2014fsa, Schuster:2014hca, Font:2013hia, Bekaert:2010hw, Bekaert:2005in, Khan:2004nj, Segal:2001di, deWit:1979sib, Fang:1978wz, Fronsdal:1978rb, Hirata:1977ss, Abbott:1976bb, Singh:1974qz, Singh:1974rc, Chakrabarti:1971rz, Yngvason:1970fy}, supersymmetric generalizations \cite{Buchbinder:2022msd, Najafizadeh:2021dsm, Najafizadeh:2019mun, Buchbinder:2019sie, Buchbinder:2019kuh, Buchbinder:2019esz}, CSP's in AdS/dS spaces \cite{Metsaev:2021zdg, Metsaev:2019opn, Metsaev:2017ytk, Metsaev:2016lhs, Buchbinder:2024jpt, Metsaev:2017myp}, field-theoretic aspects of interactions \cite{Metsaev:2018moa, Bekaert:2017xin, Metsaev:2017cuz, Bekaert:2010hp, Berends:1985xx, Berends:1984rq, Iverson:1971hq, Iverson:1970kyn}, and thermodynamics with CSP's \cite{Schuster:2024wjc}.

\section{Gravitational Waves and Worldline Formalism}\label{2}
Gravitational time-delay is the foundation of gravitational wave detection at experiments like LIGO \cite{LIGOScientific:2016aoc}, Virgo \cite{Virgo}, Kagra \cite{Kagra}, LISA \cite{LISA}, TianQin \cite{TianQin}, and a standard textbook observable (see e.g. \cite{Maggiore:2007ulw}). To make contact with the description of other interactions with $\rho_g \neq 0$, it is useful to first recast the computation of the gravitational time-delay and strain in terms of a linearized helicity-2 field with Fierz-Pauli action in flat Minkowski space, coupled to a current sourced by the worldlines of ``mirror'' and ``laser'' particles.

We start by discussing linearized gravitational waves in section \ref{2.1}, to establish notation. We introduce the matter worldlines and their coupling to gravitational waves in section \ref{2.2}, and compute the strain in section \ref{2.3}.


\subsection{Gravitational Waves and Associated Observables}\label{2.1}
We briefly introduce the conventions used throughout this paper. We work with the mostly negative metric, $\mathbf{g}_{\mn} = \text{diag}(+,-,-,-)$. Greek indices $\mu,\nu,\dots$ label spacetime indices, Latin indices $i,j,\dots$ label spatial indices, and $a$ or $(a)$ label polarization states. Dots over worldline variables refer to derivatives with respect to the worldline parameter $\tau$.

We start with the Einstein-Hilbert action,
\begin{equation}
    S = \int \dd[4]{x} \sqrt{-g} \left( \frac{1}{2 \kappa} R + \mathcal{L}_{\text{m}} \right),
\end{equation}
where $\kappa = 8 \pi G_N$. Considering perturbations about the flat Minkwoski metric $g_{\mn} = \mathbf{g}_{\mn} + h_{\mn}$, and keeping terms to $\mathcal{O}(|h_{\mn}|^2)$, we obtain the Fierz-Pauli action,
%
\begin{align}
\label{action}
    S_{\text{FP}} = \frac{1}{2 \kappa} &\int \dd[4]{x} \bigg( \frac{1}{2} \p_\rh h_{\mn} \p^\rh h^{\mn} - \p_\rh h_{\mn} \p^\n h^{\rh \m} \nonumber \\
    & \qquad \qquad - \frac{1}{2} \p_\m h^\s_\s \p^\m h^\s_\s + \p_\n h^{\mn} \p_\m h^\s_\s \bigg),
\end{align}
%
where $h^\s_\s$ is the trace of the metric, and indices are raised and lowered with Minkowski metric, $\mathbf{g}_{\mn}$. This action is invariant under the gauge transformation, 
\begin{equation}
\label{fpgauge}
    h_{\mn} \rightarrow h_{\mn} + \p_\m \chi_\n + \p_\n \chi_\m,
\end{equation}
which is just linearized diffeomorphism invariance inherited from the full non-linear theory. For a more detailed review of linearized GR, the reader is referred to \cite{Misner:1973prb, Carroll:2004st, Maggiore:2007ulw}. For radiation problems, a convenient gauge choice is the de-Donder gauge,
\begin{equation}
\label{ddgauge}
    \p^\m \left( h_{\mn} - \frac{1}{2} h^\s_\s \mathbf{g}_{\mn} \right) = 0 ,
\end{equation}
where the equation of motion reduces to 
\begin{equation}
    \Box \left( h_{\mn} - \frac{1}{2} h^\s_\s \mathbf{g}_{\mn} \right) = 0 .
\end{equation}
Defining the trace-reversed metric, $\overline{h}_{\mn} = h_{\mn} - (1/2) h^\s_\s \mathbf{g}_{\mn} $, the de-Donder gauge condition and equation of motion simplify to 
\begin{equation}
\label{eomgw}
    \p^\m \overline{h}_{\mn} = 0 \quad \mbox{ and } \quad \Box \overline{h}_{\mn} = 0
\end{equation}
respectively. This covariant gauge-fixing admits a residual freedom under \eqref{fpgauge} with $\partial_i \dot \chi =0$. This residual freedom can be used to render the polarization tensors spatially transverse ($\partial_i h_{ij}=0$) and traceless ($h_i^i=0$). For plane-wave solutions, this is the transverse-traceless (TT) gauge, which is commonly used in the gravitational wave literature. Fixing the gravitational wave to be propagating along the $z$-axis, with 4-vector $k_\m = (\om, 0, 0, \om)$, a basis of polarization tensors is 
\begin{align}
\label{pol}
    H^{(+)}_{\mn} = 
    \begin{pmatrix}
        0 & 0 & 0 & 0 \\
        0 & - 1 & 0 & 0 \\
        0 & 0 & 1 & 0 \\
        0 & 0 & 0 & 0 \\
    \end{pmatrix} \quad 
    H^{(\cross)}_{\mn} = 
    \begin{pmatrix}
        0 & 0 & 0 & 0 \\
        0 & 0 & - 1 & 0 \\
        0 & - 1 & 0 & 0 \\
        0 & 0 & 0 & 0 \\
    \end{pmatrix}
\end{align}
We will focus on monochromatic solutions to the wave equation \eqref{eomgw} propagating along the $z$ axis, given in general by 
\begin{equation}
\label{gw}
    h_{\mn}(z) = \sum_{a = +,\cross} h_{a} H^{(a)}_{\mn} \cos \left( k \cdot z + \phi^{(a)} \right),
\end{equation}
from which more general solutions can be built as superpositions. These solutions have been expressed in the detector (rest) frame. Gauge transformations that preserve the covariant de-Donder gauge condition \eqref{eomgw} but not the spatial transverse condition $\partial_i h_{ij}=0$ are equivalent to rigid Lorentz transformations of the observer's reference frame; independence of observables under these transformations is an important consistency check on computations, and is discussed further in Appendix \ref{A}. 

Gravitational waves incident on an interferometric GW detector induce opposite phase-shifts in the light reflected in the two arms, producing an experimentally detectable interference pattern (see e.g.~chapter 9 of  \cite{Maggiore:2007ulw}).  The phase acquired in each arm is closely related to the round-trip time (in the beam-splitter's reference frame) for light to
travel from the beam-splitter to the opposing mirror and back in a given GW background, $T_{\text{GR}}$, which differs from the flat-space travel time $T_{\text{M}}$.  The round-trip time is gauge-invariant even for one arm, and can be calculated from the path of a single photon. In this paper we will use $T_{\text{GR}}$ for a single arm, and the related dimensionless time-delay 
\begin{align}
\label{dtd}
    g = \frac{(T_{\text{GR}}-T_{\text{M}})}{T_M},
\end{align} 
as a proxy for the interferometer signal.

\subsection{Worldline Formalism}\label{2.2}
In this calculation, we treat the photon as a massless scalar, and the mirrors as massive scalars. The action for a relativistic free scalar particle in Minkwoski ($\mathbf{g}_{\mn}$) spacetime with some arbitrary mass $m$ is conveniently written as,
\begin{equation}
\label{freeworldline}
    S_{\text{free}} = \frac{1}{2} \int \dd{\tau} \bigg[ \frac{\gd^\m (\tau) \gd^\n (\tau)}{e(\tau)} \mathbf{g}_{\mn} + e (\tau) m^2 \bigg]
\end{equation}
Here, $\gamma_\m (\tau)$ is an \emph{arbitrary} parametrization of the worldline (geodesic) of the particle in terms of a ``time'' parameter $\tau$, $\dot\gamma_\m \equiv \partial_\tau\gamma_\m $, and $e(\tau)$ is an auxiliary ``einbein'' field. Note that 
the einbein equation of motion implies $e=\sqrt{-\dot\gamma^2}/m$, which can be substituted back into the action to recover the more familiar relativistic free-particle action 
$S_{\text{free}} = m \int \dd{\tau} \sqrt{- \gd^2}$. However, keeping the einbein field explicit yields a simpler, polynomial action in $
\dot\gamma$ and is also necessary to take the massless limit.

To couple the particle to a helicity-2 background $h_{\mu\nu}(x)$, we add the interaction term 
\begin{align}
    &S_{\text{int}} = \int \dd[4]{x} h_{\mn} (x) T^{\mn} (x), \nonumber\end{align}
    where $T^{\mn}$ is the rank-2 matter current 
\begin{align}    \label{intgw}
    T^{\mn} (x) &= \frac{1}{2}  \int \dd{\tau} \frac{\gd^\m (\tau) \gd^\n (\tau)}{e(\tau)} \del^{(4)} (x - \g(\tau))
\end{align}
identified with the energy-momentum tensor (EMT). This EMT is not conserved: $\p_\m T^{\mn} \propto \gdd^\n (\tau)$. This non-conservation is well known and physical --- only the total EMT, including the energy-momentum of all particles and/or fields driving the acceleration of $\gamma^{\mu}$, is conserved. For more details, the reader is referred to \cite{Misner:1973prb, Ortin:2015hya}. 

Using the plane wave solution in \eqref{gw}, and integrating out the delta function in the interaction term \eqref{intgw}, we obtain the action for $\gamma^{\mu}$ in a GW background at leading order in $G_N$,  
\begin{equation}
    S = \frac{1}{2} \int \dd{\tau} \bigg[ \frac{\gd^\m \gd^\n}{e} \left( \mathbf{g}_{\mn} + h_{a} H^{(a)}_{\mn} \cos \left( k \cdot \g + \phi^{(a)} \right) \right) + e m^2 \bigg].
\end{equation}
This is equivalent to \eqref{freeworldline} with $\mathbf{g}_{\mn}$ replaced by the dynamical gravitational wave metric, but we have emphasized the flat-space perspective that we will later generalize to study CSPs. 
The repeated index $a$ sums over  the two polarization states $+$ and $\times$, and the $\tau$-dependence of $\gamma$ and $\dot\gamma$ is suppressed. The resulting equation of motion (which we also refer to as the geodesic equation) is 
\begin{align}
\label{gweom}
    - &\frac{\gd^\rh \gd^\s}{e} k_{\m} h_{a} H^{(a)}_{\rh \s} \sin \left( k \cdot \g + \phi^{(a)} \right) \nonumber \\ 
    &= 2 \partial_{\tau} \bigg[ \frac{ \gd^\n}{e} \left( \mathbf{g}_{\mn} + h_{a} H^{(a)}_{\mn} \cos \left( k \cdot \g + \phi^{(a)} \right) \right) \bigg].
\end{align}
Varying the action with respect to the einbein gives the constraint
\begin{equation}
\label{gweome}
    \gd^\m \gd^\n \left( \mathbf{g}_{\mn} + h_{a} H^{(a)}_{\mn} \cos \left( k \cdot \g + \phi^{(a)} \right) \right) = e^2 m^2.
\end{equation}
Fixing the einbein to any non-zero constant, $e = \al \neq 0$, results in an $\al$-independent equation of motion (EOM) \eqref{gweom}, while the constraint equation \eqref{gweome}  scales uniformly with $\al$.

To solve the EOM and einbein constraint to linear order in $h$, it is convenient to expand the solution in $h$ as 
\begin{equation}
    \label{ansatz}
    \g_\m (\tau) = \g^{(0)}_\m (\tau) + h_a \g^{(a)}_{\m} (\tau),
\end{equation}
and plug these into the EOM \eqref{gweom} and the einbein constraint \eqref{gweome}, and expand the resulting equations in $h$. The zeroth-order equations are simply the Minkowski EOM and mass-shell constraint for $\gamma^{(0)}$. The first-order equations can be solved for $\g^{(a)}$ in terms of $\g^{(0)}$ and the background field. Explicitly, the zeroth order and linear EOMs are
\begin{align}
    \gdd^{(0)}_\m &= 0 \quad \text{and} \quad \gdd^{(a)}_\m = A^{(a)}_\m \sin \left( \kg^{(0)} + \phi^{(a)} \right)
\end{align}
where we have introduced the coefficient function
\begin{align}
    A^{(a)}_\m = &\left( \kgd^{(0)} \right) \gd^{(0)\n} H^{(a)}_{\mn} - \frac{\gd^{(0)\rh} \gd^{(0)\s}}{2} k_{\m} H^{(a)}_{\rh \s},
\end{align}
and dropped terms proportional to $\gdd^{(0)}$ which vanish by the zeroth-order EOM.  Note that the coefficient function $A^{(a)}$ is $\tau$-independent because it only depends on the worldline through $\gd^{(0)}$, which is $\tau$-independent by the zeroth-order EOM. The general solutions to these EOMs are
\begin{align}
\label{gwgeodesics}
    \g^{(0)}_\m (\tau) &= b^{(0)}_\m + c^{(0)}_\m \tau, \nonumber \\ 
    \g^{(a)}_\m (\tau) = b^{(a)}_\m + &c^{(a)}_\m \tau + \frac{A^{(a)}_\m}{\left( \kgd^{(0)} \right)^2} \sin \left( \kg^{(0)} + \phi^{(a)} \right),
\end{align}
where $b^{(0)}_\m$, $c^{(0)}_\m$, $b^{(a)}_\m$, and $c^{(a)}_\m$ are constants of integration that must be fixed by boundary conditions.  For each particle considered in a background with $a$ GW polarizations, the combination of boundary conditions and einbein constraints must fix $2(a+1)$ undetermined four-vectors or $8(a+1)$ real numbers. 
Note that $c^{(0)}_\m$ can be interpreted as the 4-momentum in Minkowski space. For a massless particle the resulting geodesics match \cite{Grishchuk:1974jy}, if one uses $c^{(0)}_\m = \Omega (1, v_x, v_y, v_z)$.

Similar to the equation of motion, the einbein constraint can also be expanded order by order:
\begin{align}
\label{gweomeexp0}
    \left(\gd^{(0)}\right)^2 &= e^2 m^2 \\
\label{gweomeexp1}
    2 \left( \gd^{(0)} \cdot \gd^{(a)} \right) + \gd^{(0)\m} \gd^{(0)\n} &H^{(a)}_{\mn} \cos \left( \kg^{(0)} + \phi^{(a)} \right) = 0
\end{align}
The zeroth order term is, as noted earlier, the mass-shell condition in Minkowski space, and the linear order term is an additional constraint when matching boundary conditions.

\subsection{Computing the Strain} \label{2.3}
\begin{figure}[ht]
\begin{center}
\includegraphics[width=85mm]{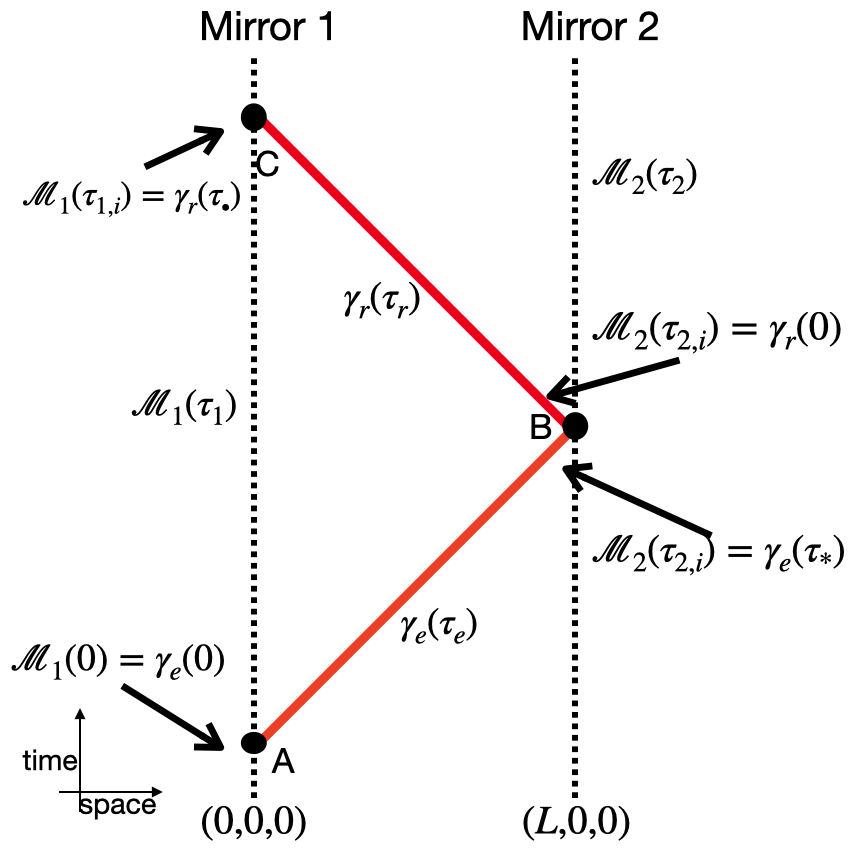}
\caption{The dotted lines correspond to the mirror geodesics, and the red lines correspond to the photon geodesics. The blobs correspond to when the photon intersects a mirror. The point A corresponds to the emission of the photon from the first mirror, the point B corresponds to the reflection of the photon at the second mirror, and point C corresponds to the photon hitting the first mirror. The difference in time coordinates between A and C is the time $T_{\text{GR}}$.}
\label{setup}
\end{center}
\end{figure}

The strain is computed in the following toy model where we consider a photon reflected between two mirrors, as can be seen in figure \ref{setup}. The mirrors are considered to be inertial point masses separated along the $x$-axis, and the gravitational wave is propagating along the $z$-axis. In this configuration, only the $+$ polarization state contributes to the signal, so we restrict our attention to it. In the counting above, we therefore have $a=1$ and a total of four 4-vectors (16 real constants) to fix for each body's trajectory.

Massive point particles not feel the passage of gravitational waves. Therefore, we are free to fix the spatial components of the mirror worldlines. We pick mirrors at the origin, and some distance $L$ along the x-axis in the detector frame. With this choice, the mirror worldlines are, 
\begin{equation}
\label{mirror}
    \mathcal{M}_1^\m (\tau_1) = (\tau_1, 0, 0, 0) \qquad \mathcal{M}_2^\m (\tau_2) = (\tau_2, L, 0, 0),
\end{equation}
where $\tau_1$, and $\tau_2$ refers to the proper time in the frame of the respective mirror. We can compute the time elapsed between events on a mirror by computing the difference in the time coordinate of the mirror between the two events.  We note that the setting of boundary conditions becomes more subtle in the presence of gauge terms as considered in Appendix \ref{A}, which induce oscillatory motion of the mirrors. We can still impose boundary conditions on the mirror's time-averaged position and velocity in this case.  Although each geodesic acquires gauge-dependent corrections, these cancel out of the final time-delay computed below, as expected.   

We turn now to the boundary conditions for the photon, which are fixed by matching to the mirrors. 
The emitted photon $\g_e (\tau_e)$ leaves $\mathcal{M}_1 (\tau_1)$ and is reflected by $\mathcal{M}_2 (\tau_2)$. The matching conditions are $\mathcal{M}_1^\m (\tau_1 = 0) = \g_e^\m (\tau_e = 0)$ (point A in figure \ref{setup}), and $\mathcal{M}_2^\m (\tau_2 = \tau_{2,i}) = \g_e^\m (\tau_e = \tau_*)$ (point B in figure \ref{setup}). More explicitly, 
\begin{align}
    \label{eat1}
    &\mathcal{M}_1 \rightarrow  \g^{(0)}_{e\m} (0) = (0,0,0,0) \text{ , } \g^{(+)}_{e\m} (0) = (0,0,0,0) \quad  \\
    \label{eat2}
    &\mathcal{M}_2 \rightarrow \g^{(0)}_{e\m} (\tau_{*}) =  (\tau_{2,i}^{(0)},L,0,0) \text{ , } \g^{(+)}_{e\m} (\tau_*) = (\tau_{2,i}^{(+)},0,0,0) 
\end{align}
Here we have expanded $\tau_{2,i} = \tau_{2,i}^{(0)} + h_+ \tau_{2,i}^{(+)}$, where $\tau_{2,i}^{(0)}$ is the leading order solution, and $\tau_{2,i}^{(+)}$ is the first order correction from the perturbative calculation. The zeroth order einbein constraint, \eqref{gweomeexp0} fixes $\tau_{2,i}^{(0)} = L/\tau_*$. To determine $\tau_{2,i}^{(+)}$, we integrate the linear order einbein constraint \eqref{gweomeexp1} over the path of the photon from $\tau = 0$ to $\tau_*$,
\begin{equation}
     \gd^{(0)}_e \cdot \g^{(+)}_e (\tau_*) = - \frac{\gd^{(0)\m}_e \gd^{(0)\n}_e H^{(+)}_{\mn}}{2 \left( \kgd^{(0)}_e \right)} \sin \left( \kg^{(0)}_e (\tau_*) \right).
\end{equation}
In computing the integral, we have used that $\g^{(0)} (\tau)$ is a linear function of $\tau$. From \eqref{eat2}, $\g^{(+)}_{e,0} (\tau_*) = \tau_{2,i}^{(+)}$, and we can use the integrated linear einbein constraint to extract its value
\begin{equation}
    \tau_{2,i}^{(+)} = \frac{\sin \left(L \omega + \phi^{(+)} \right) - \sin \left( \phi^{(+)} \right)}{2 \omega}.
\end{equation}
Integrating the einbein constraint is a useful shortcut that allows the direct inference of the time delay, without explicitly computing the geodesics. However, for the sake of completeness (and as a check on the calculation), geodesics with appropriate boundary conditions are listed in appendix \ref{B}.

Similarly, the reflected photon trajectory $\g_r (\tau_r)$ satisfies $\mathcal{M}_2^\m (\tau_2 = \tau_{2,i}) = \g_r^\m (\tau_r = 0)$ (point B in figure \ref{setup}), and $\mathcal{M}_1^\m (\tau_1 = \tau_{1,i}) = \g_r^\m (\tau_e = \tau_{\bullet})$ (point C in figure \ref{setup}). An analogous treatment of  boundary conditions for the reflected photon determines that
\begin{align}
\label{gwstrain}
    &\tau_{1,i} = 2L + \Delta_{\text{GR}} \cos \left( L\omega + \phi^{(+)} \right), \\
    &\text{ where }  \Delta_{\text{GR}} = \frac{ h_{+} \sin \left(L \omega \right)}{2 \omega}. \nonumber
\end{align}
Considering the geodesics of $\mathcal{M}_1$ in \eqref{mirror}, we see that $\tau_{1,i}$ is the value of the time geodesic when the photon hits it after being reflected -- this is the total time in the detector frame for the photon to complete one round trip in the presence of the GW background -- $\tau_{1,i} = T_{\text{GR}}$ as defined in \eqref{dtd}. Following the same, and using that $T_{\text{M}} = 2L$, we get,
\begin{equation}
    g_{\text{GR}} = \frac{\Delta_{\text{GR}}}{2L} \cos \left( L\omega + \phi^{(+)} \right).
\end{equation}
This closely matches the result in chapter 9 of \cite{Maggiore:2007ulw}. 

For photons emitted at a later time $t$, we replace $\phi^{(+)} \rightarrow \phi^{(+)} + \om t$, to account for the phase evolution of the wave. The overall factor of $\Delta_{\text{GR}}$, is the amplitude that controls the size of the effect.


\section{CSP Gravitational Waves}\label{3}
In this section, we generalize the approach of section \ref{2} to compute the strain for a GW-CSP background. We start by introducing the $\e$-space formalism of \cite{Schuster:2013pxj, Schuster:2013vpr, Schuster:2013pta, Schuster:2014hca,Schuster:2023xqa,Schuster:2023jgc}, which will allow us to generalize the scalar matter worldline interactions with a rank-2 tensor field (as in linearized GR) to the case of a CSP field. 
In section \ref{newCSPIntro} we summarize key results from \cite{Schuster:2023xqa} on CSP gauge theory and $\eta$-space integration that we will use directly in this work. For context and derivations, we refer the reader to Sections II and III and Appendix A of \cite{Schuster:2023xqa}.
In section \ref{etarho0} we recast to $\e$-space the $\rho_g=0$ interaction term used in section \ref{2}. We then generalize this interaction, at leading order in $G_N$, to non-zero $\rho_g$ in section \ref{gravrho}, and extend the calculation of strain from section \ref{2} to non-zero $\rho_g$ in  section \ref{strainrho},

\subsection{Results for Continuous Spin Gauge Theories}
\label{newCSPIntro}
A gauge field theory for CSPs was developed in \cite{Schuster:2014hca} and subsequently coupled to matter particles in \cite{Schuster:2023xqa}. 
This theory is based on a (bosonic) CSP field $\Psi(\eta^\mu,x^\mu)$, where $x^\mu$ is the usual space-time coordinate and the dependence of $\Psi$ on the new four-vector coordinate $\eta^\mu$ encodes the particle's spin (with substantial gauge redundancy). Because we are dealing here with the response of matter particles to a CSP plane wave satisfying the vacuum equations of motion (i.e. on-shell), we will need only four key results: the mode expansion of the CSP field, its coupling to a current, evaluation of certain integrals over $\eta$, and the embedding of on-shell (linearized) gravitational waves and the stress-energy tensor in the $\eta$-space action with $\rho=0$. In the interest of brevity, we state the relevant results with minimal commentary, referring the reader to specific sections of \cite{Schuster:2023xqa} for elaboration and context. 

\paragraph{Field Mode Expansion:} 
Up to gauge terms, a general solution to the CSP equation of motion can be decomposed as 
\begin{equation}
\label{modeexp}
    \Psi (\e, x) = \int \frac{\dd[3]{\vb{k}}}{(2 \pi)^3} \frac{1}{2|\vb{k}|}  \sum_{h} \bigg( a_h (\vb{k}) \psi_{h,\vb{k}} e^{-i \vb{k}\cdot x }  + \text{c.c.} \bigg) \bigg|_{k^0 = |\vb{k}|}
\end{equation}
with 
\begin{align}
\label{CSPModes}
    \psi_{h,\vb{k}} = e^{- i \rh_g \eq} \times \Biggl\{ 
    \begin{array}{cc}
        (i \e \cdot \ep^{+})^h & h \geq 0 \\
        (-i \e \cdot \ep^{-})^{-h} & h \leq 0
    \end{array}
\end{align} 
(see (2.30) and (2.31) of \cite{Schuster:2023xqa}). This decomposition is given in terms of a set of null ``frame'' vectors $\ep^\pm,q$ for given momentum $k$ satisfying $\epsilon^-={\epsilon^+}^*$, $k\cdot q=1$, and $\epsilon_+\cdot\epsilon_- = -2$ with all other inner products vanishing. As with the choice of polarization vectors in more familiar gauge theories, expansions using different choices of $\epsilon^\pm$ and $q$ are related by a gauge + little group transformation.  For a wave propagating along the $z$-axis, $k_\m = (\om, 0, 0, \om)$, a convenient choice is $\ep^{\pm}_{\m} = (0, 1, \pm i, 0)$ and $q_\m = (1/2\om, 0, 0, -1/2\om)$, for which the modes $\psi_{h,\vb{k}}$ are eigenmodes of the usual helicity operator ${\mathbf{J}}\cdot { \mathbf{\hat k}}$ with eigenvalue $h$.  In the next subsection, we will relate the previously considered $H^{(+)}$ and $H^{(\cross)}$ modes to real superpositions of the $h=\pm 2$ modes with $\rho_g = 0$. 

\paragraph{CSP Field Coupling to Currents:}
A general linear CSP coupling to an external current can be written as
\begin{equation}
\label{cspint}
    S_{\text{int}} = \int [\dd[4]{\e}] \dd[4]{x} \del' (\e^2 + 1) J(\e,x) \Psi(\e,x),
\end{equation}
where $J(\e,x)$ must respect the continuity condition
\begin{equation}
\label{ContinuityCondition}
    \delta (\e^2 + 1) ( \p_x \cdot \p_\e + \rh_g ) J(\e,x) = 0
\end{equation}
in order to ensure gauge invariance (see (3.17) and (3.23) of \cite{Schuster:2023xqa}).  
In the above, $[\dd[4]{\e}]$ is a regulated measure discussed briefly below and in depth in Appendix A of \cite{Schuster:2023xqa}, which satisfies the usual integration-by-parts and $\delta$-function identities; $\delta'(a)=\frac{d}{da}\delta(a)$; and $\p_x$ and $\p_\e$ denote $\partial/\partial_x$ and $\partial/\partial_\eta$, respectively.   

We also refer the reader to Section III of \cite{Schuster:2023xqa} for a pedagogical overview of currents sourced by spinless particles that give rise to scalar- and vector-like interactions with CSP fields (those dominated at energies $\gg\rho_g$ by the helicity 0 or $\pm 1$ modes of the CSP), which we generalize here to the tensor-like case (dominated by helicity 2).

We also note here an important convention difference from \cite{Schuster:2023xqa} to the present work. The analysis of \cite{Schuster:2023xqa} used  canonically normalized (mass-dimension 1) fields $\Psi(\eta,x)$ and currents $J(\eta,x)$ of mass-dimension 3. To mimic more closely the conventions of linearized GR, we will work from here forward with fields and currents that have been scaled by $\sqrt{\kappa}$ and $1/\sqrt{\kappa}$ respectively (where $\kappa = 8\pi G_N$), so that $\Psi$ is dimensionless and currents have mass-dimension 4. 

\paragraph{Integration over $\e$:}
Integrals $\int d^4\eta \delta'(\eta^2+1) f(\eta)$ are naively divergent; loosely speaking, the notation $[d^4\eta]$ signifies normalizing by this divergent factor --- a procedure that can be made rigorous using generating function arguments or by analytic continuation of $\eta^0$, which renders the surface of integration $\eta^2+1 =0$ compact. Both approaches are discussed in Appendix B of \cite{Schuster:2023xqa}.  All integrals used here are of the form 
\begin{align}
    \int [d^4\eta] \delta'(\eta^2+1) \eta^{\mu_1}\dots\eta^{\mu_n} e^{i\eta\cdot V}
\end{align}
with $n=0$ or 2, or its $V\to 0$ limit with $n=0$, 2, or 4. All such integrals follow from repeated differentiation with respect to $V$ of (A18) from \cite{Schuster:2023xqa},
\begin{align}
    \int [d^4\eta] \delta'(\eta^2+1) e^{i\eta\cdot V} = J_0(\sqrt{-V^2}).
    \label{etaInt0rho}
\end{align}
For example,
\begin{align}
    \int &[d^4\eta] \delta'(\eta^2+1) \eta^\mu \eta^\nu  e^{i\eta\cdot V} \nonumber \\
    & = -\frac{\partial}{\partial V_\mu} \frac{\partial}{\partial V_\nu} J_0(\sqrt{-V^2}) \nonumber \\
    & = - {\mathbf{g}}^{\mu\nu}\frac{J_1(\sqrt{-V^2})}{\sqrt{-V^2}}  + V^\mu V^\nu \frac{J_2(\sqrt{-V^2})}{V^2} 
    \label{etaInt2rho}
\end{align}
and from the $V\to0$ limits of \eqref{etaInt0rho} and \eqref{etaInt2rho},
\begin{align}
\int [d^4\eta] \delta'(\eta^2+1)&=  1,\\ 
\int [d^4\eta] \eta^\mu \eta^\nu \delta'(\eta^2+1)&=  -\frac{1}{2} {\mathbf{g}}^{\mu\nu}.\label{etaInt02}
\end{align}
Taking two more derivatives and then the $V\to0$ limit we obtain 
\begin{align}
    \int &[d^4\eta] \delta'(\eta^2+1) \eta^\mu \eta^\nu  \eta^\rho\eta^\sigma \nonumber \\
    &= \frac{1}{8} \left({\mathbf{g}}^{\mu\nu} {\mathbf{g}}^{\rho\sigma} + {\mathbf{g}}^{\mu\rho} {\mathbf{g}}^{\nu\sigma} +  {\mathbf{g}}^{\mu\sigma}{\mathbf{g}}^{\nu\rho} \right).
    \label{etaInt4}
\end{align}

\paragraph{Extracting the Graviton Mode at $\rho_g = 0$:}
When $\rho_g =0$, the free action of \cite{Schuster:2023xqa} can be decomposed into a sum of familiar actions for rank-$h$ tensor fields encoding helicities $\pm h$ (Section IIB of \cite{Schuster:2023xqa}), and the interaction \eqref{cspint} likewise decomposes into a sum of couplings between helicity-$h$ modes and conserved rank-$h$ tensor currents (Eqs.~(3.18)-(3.21) and Appendix B1 of \cite{Schuster:2023xqa}).
We need only two results from this general machinery: First,  $h=\pm 2$ gravitational wave modes are embedded in $\Psi$ as 
\begin{align}
\label{etagraviton}
\Psi(\eta,x) = 2 \eta^\mu \eta^\nu h_{\mu\nu}(x) 
\end{align}
(eq.~(2.8) of \cite{Schuster:2023xqa} with trace terms omitted since they vanish on the support of vacuum equations of motion). 

Second, a current of the form
\begin{align}
\label{0RhoEtaCurrentAnsatz}
J(\eta,x) = \left(\eta^{\mu}\eta^{\nu} + \frac{1}{2} {\vb{g}}^{\mu\nu} \right) T_{\mu\nu}
\end{align}
sources the graviton field exclusively --- this is eq.~(3.19) of \cite{Schuster:2023xqa} up to the $\kappa$-rescaling noted above.

\subsection{Linearized GR Modes in $\e$-space}
\label{etarho0}
Using the results above, we now make contact between the formulation of linearized GR coupled to worldlines in Sec.~\ref{2} and the $\e$-space formalism of \cite{Schuster:2014hca,Schuster:2023xqa} with $\rho=0$.

\paragraph{Mode Expansion for $\rho_g = 0$:} 
In the $\rho_g \to 0$ limit, the $h=\pm 2$ modes from \eqref{CSPModes} are monomials quadratic in $\eta$,
\begin{align}
\label{CSPModesH2}
    \psi_{\pm 2,{\mathbf{k}}} = (i \e \cdot \ep^{\pm})^2.
\end{align} 
Note that the polarization tensors defined in \eqref{pol} can be written as  
\begin{equation*}
     H^{(+)}_{\mn} = \frac{ - \left( \ep^{+}_{\n} \ep^{+}_{\n} + \ep^{-}_{\m} \ep^{-}_{\n} \right)}{2},  \quad H^{(\cross)}_{\mn} = \frac{i \left( \ep^{+}_{\n} \ep^{+}_{\n} - \ep^{-}_{\m} \ep^{-}_{\n} \right)}{2},
\end{equation*}
so that 
\begin{align}
\psi_{+2,\mathbf{k}}+ \psi_{-2,\mathbf{k}} &= (2\eta^\mu\eta^\nu) H^{(+)}_{\mu\nu} \\ 
-i (\psi_{+2,\mathbf{k}}- \psi_{-2,\mathbf{k}}) &= (2\eta^\mu\eta^\nu) H^{(\cross)}_{\mu\nu}.
\end{align}

Thus, taking mode amplitudes $a_{2} (\vb{k}) = a_{-2} (\vb{k}) = (2 \pi)^3 2 |\vb{k}| h_+ \del^{(3)} (\vb{k} - \om \zh)$ and similarly for $-\vb{k}$ to obtain a real solution gives 
\begin{equation}
\label{0RhoCSP}
    \Psi (\e, x) = (2 \eta^\mu \eta^\nu) h_+ \cos \left(\kx + \phi^{(+)} \right) H^{(+)}_{\mu\nu}.
\end{equation}
This is the $\eta-$space form of the field that corresponds to the GW background of section 2, with the $(2\eta^\mu\eta^\nu)$ factor as expected from \eqref{etagraviton}. 

\paragraph{Worldline Current at $\rho_g = 0$:} 
Following \eqref{0RhoEtaCurrentAnsatz}, the gravitational current at $\rho_g = 0$ is
\begin{equation*}
    J(\e, x) = \left(\e^\m \e^\n + \frac{1}{2}\mathbf{g}^{\mn} \right) T_{\mn}.
\end{equation*}
The continuity condition \eqref{ContinuityCondition}  on this ansatz reduces to $ \delta (\e^2 + 1)  \e^\m \p^\n T_{\mn} = 0$, which is equivalent to the usual conservation law $\partial_\nu T_{\mu\nu} =0$. Using the definition of $T_{\mn}$ from \eqref{intgw}, we have
\begin{equation}
\label{0RhoEtaCurrent}
    J(\e, x) = \int \dd{\tau} \left( 2\e^\m \e^\n + \mathbf{g}^{\mn}\right) \frac{\gd_\m (\tau) \gd_\n (\tau)}{4} \delta^{(4)} (x_\al - \g_\al (\tau)) 
\end{equation} 

\paragraph{Interaction Term:} We can now combine \eqref{0RhoEtaCurrent} and \eqref{0RhoCSP} to derive a matter action in the presence of a gravitational wave background. We have 
\begin{widetext}
\begin{equation}
    S_{\text{int}} = \frac{h_{+}}{2} \int \dd{\tau} [\dd[4]{\e}] \del' (\e^2 + 1) \frac{\dd[4]{l}}{(2 \pi)^4} \dd[4]{x} \left( \left(i \e \cdot \ep^{+} \right)^2 + \left(i \e \cdot \ep^{-} \right)^2 \right) \left( 2 (\egd)^2 + \gd^2 \right) \cos \left(\kx + \phi^{(+)} \right) e^{-i l (x - \g )}  
\end{equation}
\end{widetext}
Integrating over $l$, $x$, and $\eta$, using  \eqref{etaInt02} and \eqref{etaInt4} for the latter, we find
\begin{align}
\label{0RhoInt}
    S_{\text{int}} &= \frac{h_{+}}{2} \int \dd{\tau} \frac{- \left( \ep^{+}_{\n} \ep^{+}_{\n} + \ep^{-}_{\m} \ep^{-}_{\n} \right)}{2} \frac{\gd^\m \gd^\n}{e} \cos
    \left(k \cdot \g + \phi^{(+)} \right)  \nonumber \\
    &= \frac{h_{+}}{2} \int \dd{\tau} H^{(+)}_{\mn} \frac{\gd^\m \gd^\n}{e} \cos
    \left(k \cdot \g + \phi^{(+)} \right)
\end{align}
This matches \eqref{intgw}, with the delta function integrated out and the $+$ polarization of the gravitational wave substituted into the action. 

\subsection{Gravity at $\rh_g \neq 0$ at leading order in $G_{N}$}
\label{gravrho}
Now, we express the linearized in gravity CSP field, and the matter current with CSP interactions in $\e$-space. For the rest of this paper, we drop the $g$ subscript in $\rh_g$. 

\paragraph{Current:} The current is determined by the $\rh$-dependent continuity condition. It is easier to work in momentum space, 
\begin{equation}
    J(\eta,x) = \int \frac{d^4 l}{(2\pi)^4} d\tau e^{-il\cdot(x-\gamma)}j(\eta, l, \dot\gamma)
\end{equation}
where the continuity condition \eqref{ContinuityCondition} becomes 
\begin{equation}
    e^{il\cdot\gamma}\delta (\e^2 + 1) (- i l \cdot \p_\e + \rh) j(\e,l,\gd) = 0 + \partial_\tau(\dots).\label{pspaceCont}
\end{equation}
Note that the continuity condition on $j(\eta,l,\dot{\gamma})$ need only be satisfied up to a total $\tau$ derivative. We also impose the boundary condition that in the $\rh \rightarrow 0$ limit, we recover \eqref{0RhoEtaCurrent}, the EMT. As discussed in \cite{Schuster:2023xqa}, this is one of the three types of $\rho\rightarrow 0$ limits that currents can obey in general, the other two corresponding to a vector-like interactions and the other scalar-like. A solution that matches this boundary condition and satisfies the continuity condition \eqref{pspaceCont} is
\begin{widetext}
\begin{equation}
\label{current}
    j(\e,l,\gd) = - \bigg( \frac{\lgd}{\rh} \bigg)^2 \left( \exp \left( - i \rh \frac{\egd}{\lgd} \right) - 1 + i \rh \frac{\egd}{\lgd} \right) + \frac{\gd^2}{2} \exp \left( - i \rh \frac{\egd}{\lgd} \right) 
\end{equation}
\end{widetext}

The exponential pieces satisfy \eqref{pspaceCont} exactly, with no total $\tau$-derivative correction.  The subtraction terms constant and linear in $\eta$ do not, but can be expressed via integration by parts in terms of total $\tau$ derivatives (which are allowed by the continuity condition ), and terms that depend on $\gdd$. The latter are precise counterparts to the non-conserved terms in the matter EMT in GR, and are higher order in $G_N$.  We note that other solutions to \eqref{pspaceCont} exist (e.g.~ forms~with different structures in the exponential) but all solutions with the same $\rho\rightarrow 0$ limiting behavior will yield the same results as \eqref{current} for observables related to plane-waves (like the dimensionless time-delay). This is demonstrated in \cite{Schuster:2023xqa}, and is absolutely crucial to allowing unambiguous predictions of the $\rho$-dependent physics. 

\paragraph{Tensor-Like CSP:} Starting from the mode expansion in \eqref{modeexp}, keeping the $\rh$ dependent pieces, and using the amplitudes defined in \eqref{0RhoCSP} to do the integral over the Fourier modes, we obtain, 
\begin{align}
\label{CSP}
    \Psi_{+} (\e, x) &= h_{+} \left( \left(i \e \cdot \ep^{+} \right)^2 + \left(i \e \cdot \ep^{-} \right)^2 \right) \nonumber \\
    & \qquad \times \left( e^{-i \rh \eq} e^{-i \left(\kx + \phi^{(+)} \right)} + e^{i \rh \eq} e^{i \left(\kx + \phi^{(+)} \right)} \right). 
\end{align}

\paragraph{Interaction Term:} The interaction term is constructed by using the $\rh$-dependent current \eqref{current}, and the CSP mode expansion \eqref{CSP},
\begin{widetext}
\begin{align}
\label{intint}
    S_{\text{int}} &= h_{+} \int \dd{\tau} [\dd[4]{\e}] \del' (\e^2 + 1)  \frac{\dd[4]{l}}{(2 \pi)^4} \dd[4]{x} \left( (i \e \cdot \ep^{+})^2 + (i \e \cdot \ep^{-})^2 \right) \bigg( \frac{\gd^2}{2} - \left( \frac{\lgd}{\rh} \right)^2 \bigg) \nonumber \\
    & \qquad \qquad \qquad \qquad \times \left( e^{-i \rh \eq} e^{-i \left(\kx + \phi^{(+)} \right)} + e^{i \rh \eq} e^{i \left(\kx + \phi^{(+)} \right)} \right)   e^{ - i \rh \frac{\egd}{\lgd} } e^{-i l (x - \g )}.
\end{align}
\end{widetext}

In this expression, we have dropped the subtraction terms in \eqref{CSP}, as they do not contribute to the equations of motion. Gauge terms can also be added, and like in the GR case, they do change the equations of motion, but they do not affect the strain or other gauge invariant observables. To simplify $\eta$-integrals, we define
\begin{equation}
\label{V}
    V^\m = \rh \left(q^\m - \frac{\gd^\m}{ \kgd} \right).
\end{equation}
Using the above definition in \eqref{intint}, and doing the integral over the Fourier modes, spacetime, and $\e$, we obtain,
\begin{align}
    S_{\text{int}} &= 2 h_{+} \int \frac{\dd{\tau}}{e} \left( \frac{\gd^2}{2} - \left( \frac{\kgd}{\rh} \right)^{2} \right) \cos \left(k \cdot \g + \phi^{(+)} \right) \nonumber \\
    & \qquad \qquad \times J_2 \left( \sqrt{-V^{2}}\right) \left( \left( \frac{\ep^{+} \cdot V}{|\ep^{+} \cdot V|} \right)^2 + \left( \frac{\ep^{-} \cdot V}{|\ep^{-} \cdot V|} \right)^2 \right).
\end{align}
Using the identities from appendix \ref{C.1}, and defining $\z = \ep^{+} \cdot \gd$, and $\zb = \ep^{-} \cdot \gd$, the above interaction term can be simplified to 
\begin{align}
\label{tensorcspint}
    S_{\text{int}} &= 2 h_{+} \int \frac{\dd{\tau}}{e} \left( \frac{\gd^2}{2} - \left( \frac{\kgd}{\rh} \right)^{2} \right) \left( \frac{ \z^2 + \zb^2}{ \z \zb} \right) \nonumber \\
    & \qquad \qquad \times J_2 \left( \left( \frac{\rh}{\kgd} \right) \sqrt{ \z \zb }\right) \cos \left(k \cdot \g + \phi^{(+)} \right).
\end{align}
Note that in the $\rh \rightarrow 0$ limit,
\begin{align}
    & \quad \frac{h_{+}}{4} \int \frac{\dd{\tau}}{e} \left( \frac{\gd^2}{2} - \left( \frac{\kgd}{\rh} \right)^{2} \right) \left( \frac{ \z^2 + \zb^2}{ \z \zb} \right) \nonumber \\
    & \qquad \qquad \times \left( \frac{\rh}{\kgd} \right)^2 \left( \z \zb \right) \cos \left(k \cdot \g + \phi^{(+)} \right)\nonumber \\ 
    &= - \frac{h_{+}}{4} \int \frac{\dd{\tau}}{e} \left( \z^2 + \zb^2 \right) \cos \left(k \cdot \g + \phi^{(+)} \right) \nonumber \\
    &= \frac{h_{+}}{2} \int \dd{\tau} \frac{\gd^\m \gd^\n}{e} H^{(+)}_{\mn} \cos \left(k \cdot \g + \phi^{(+)} \right),
\end{align}
and we nicely recover \eqref{0RhoInt} (and therefore, \eqref{intgw}) for the GR interaction term in a GW background.

\subsection{Matter Trajectories in a $\rh_g \neq 0$ Gravitational Wave Background, and the Strain} 
\label{strainrho}
In this subsection, we compute the equation of motion with the interaction term \eqref{tensorcspint}. We then match boundary conditions on the geodesics that satisfy the equation of motion, to extract the strain signature.

\paragraph{Equation of Motion:} The interaction term in the Lagrangian for a background CSP field interacting with matter in the worldline formalism is
\begin{align}
    L_{\text{int}} &= \frac{2 h_{+}}{e} \left( \frac{\gd^2}{2} - \left( \frac{\kgd}{\rh} \right)^{2} \right) \left( \frac{ \z^2 + \zb^2}{ \z \zb} \right) \nonumber \\
    & \qquad \qquad \times J_2 \left( \left( \frac{\rh}{\kgd} \right) \sqrt{ \z \zb }\right) \cos \left(k \cdot \g + \phi^{(+)} \right).
\end{align}
Once again, we expand the worldline $\gamma$ in $h$ using \eqref{ansatz}, to obtain zeroth and linear-order equations of motion
\begin{equation}
\label{cspeom}
    \gdd^{(0)}_\m = 0 \quad \text{and} \quad \gdd^{(+)}_\m = A^{(+)}_\m \sin \left(\kg^{(0)} + \phi^{(+)} \right).
\end{equation}
These are identical to \eqref{gweom}, except that the coefficient function $A^{(+)}_\m$, given in Appendix \ref{C.2}, is now deformed by $\rh$. However, as in the GW case, the coefficient function is $\tau$-independent. 

The solution to the zeroth and linear order equation of motion are 
\begin{align}
\label{cspgeodesics}
    &\g^{(0)}_\m (\tau) = b^{(0)}_\m + c^{(0)}_\m \tau, \nonumber \\
    \g^{(+)}_\m (\tau) = b^{(+)}_\m &+ c^{(+)}_\m \tau + \frac{A^{(+)}_\m}{(\kgd^{(0)})^2} \sin \left(\kg^{(0)} + \phi^{(+)} \right).
\end{align}
The constants of integration are the same as in \eqref{gwgeodesics}, so the boundary condition matching procedure is no different that what we did in section \ref{2.3}.

The einbein constraint in the presence of the background CSP field is 
\begin{align}
\label{cspeome}
    \frac{\gd^2}{2} &= - 2 h_{+}\left( \frac{\gd^2}{2} - \left( \frac{\kgd}{\rh} \right)^{2} \right) \left( \frac{ \z^2 + \zb^2}{ \z \zb} \right) \nonumber \\
    & \qquad \qquad J_2 \left( \left( \frac{\rh}{\kgd} \right) \sqrt{ \z \zb }\right) \cos \left(\kg  + \phi^{(+)} \right) + e^2 m^2
\end{align}
Expanding this constraint, we get the mass-shell condition from the zeroth order constraint, as noted in \eqref{gweomeexp0}. The constraint linear in $h_+$ is 
\begin{widetext}
\begin{equation}
\label{cspeomeexp1}
    \gd^{(0)} \cdot \gd^{(+)} = - 2 h_{+}\left( \frac{\left( \gd^{(0)} \right)^2}{2} - \left( \frac{\kgd^{(0)}}{\rh} \right)^{2} \right) \left( \frac{\ep^{+} \cdot \gd^{(0)}}{\ep^{-} \cdot \gd^{(0)}} + \frac{\ep^{-} \cdot \gd^{(0)}}{\ep^{+} \cdot \gd^{(0)}}  \right) J_2 \left( \frac{\rh}{\kgd^{(0)}} \sqrt{ \left( \ep^{+} \cdot \gd^{(0)} \right) \left( \ep^{-} \cdot \gd^{(0)} \right) }\right) \cos \left(\kg^{(0)} + \phi^{(+)} \right) 
\end{equation}
\end{widetext}

\paragraph{Strain:} For the mirror geodesics, we once again pick $e(\tau) = 1/m$. The zeroth order geodesics are fixed using the same arguments as in section \ref{2.3}. Using these, we find that the $\rh$ deformed $A^{(+)}_\m$ is vanishing for the mirrors, since the argument of the Bessel function is vanishing. Furthermore, the linear order geodesic constraint also reduces to $\gd^{(0)} \cdot \gd^{(+)} = 0$ for the same reason. Therefore, in the presence of CSP's, the mirror geodesics are still the same as in \eqref{mirror}.

For the emitted photon $\g_e (\tau_e)$, the boundary conditions that we match are exactly the same as in \eqref{eat1}, and \eqref{eat2}. Using the zeroth order einbein constraint for the CSP case we again get that $\beta^{(0)}_e = L$. Like we did for the GW case, $\beta_e^{(+)}$ can be extracted by integrating the linear order einbein constraint over the path of the photon.
\begin{align}
    \gd^{(0)}_e \cdot \gd^{(+)}_e (\tau_*) &= \frac{4 h_{+}}{\rh^2} \left( \kgd^{(0)}_e \right)  J_2 \left( \frac{\rh}{\om} \right) \nonumber \\
    & \qquad \times \sin \left(\kg^{(0)} (\tau_*) + \phi^{(+)} \right) \label{simplifiedEinbeinRho}
\end{align}
Where once again, we have utilized that $\g^{(0)}_\m$ is a linear function of $\tau$. Using the values of the coefficients from \eqref{eat1}, and \eqref{eat2}, we have 
\begin{equation}
    \beta^{(+)}_e = 4 h_{+} \left( \sin \left(L \omega + \phi^{(+)} \right) - \sin \left( \phi^{(+)} \right) \right)\left( \frac{\om}{\rh^2} \right) J_2 \left( \frac{\rh}{\om} \right) 
\end{equation}

The constants for the reflected photon $\g_r (\tau_r)$ are determined in a similar fashion. An analogous implementation of the integrated linear order einbein constraint fixes 
\begin{align}
    &\tau_{1,i} = 2L + \Delta_{\text{CSP}} \cos \left(L\om + \phi^{(+)} \right), \\
    &\text{ where }  \Delta_{\text{CSP}} = 8 h_{+} \sin (L\om) \left( \frac{\om}{\rh^2} \right) J_2 \left( \frac{\rh}{\om} \right). \nonumber
\end{align}
Noting that $\tau_{1,i} = T_{\text{CSP}}$, the time taken in the detector frame for a photon to complete a round trip, and recalling that $T_{\text{M}} = 2L$, and using \eqref{dtd}, we have the dimensionless time-delay in the presence of a CSP gravitational wave background,
\begin{equation}
\label{cspstrain}
    g_{\text{CSP}} = \frac{\Delta_{\text{CSP}}}{2L} \cos \left(L\om + \phi^{(+)} \right).
\end{equation}
The phase structure of this signal is identical to \eqref{gwstrain}. The overall amplitude is $\Delta_{\text{CSP}}$ which is deformed by $\rho$, and encodes the changes to the strain due to the CSP nature of gravity considered here.

As a check, we consider the $\rh/\om \ll 1$ limit, where the strain becomes,
\begin{equation}
\label{cspstrainexp}
    \Delta_{\text{CSP}} =  \frac{h_{+} \sin (L\om)}{\om} \left( 1 - \frac{1}{12} \left( \frac{\rh}{\om} \right)^2 + \mathcal{O} \left( \frac{\rh}{\om} \right)^4 \right).
\end{equation}
Clearly, setting $\rh = 0$ recovers \eqref{gwstrain}.

\begin{figure}[h]
\begin{center}
\includegraphics[width=85mm]{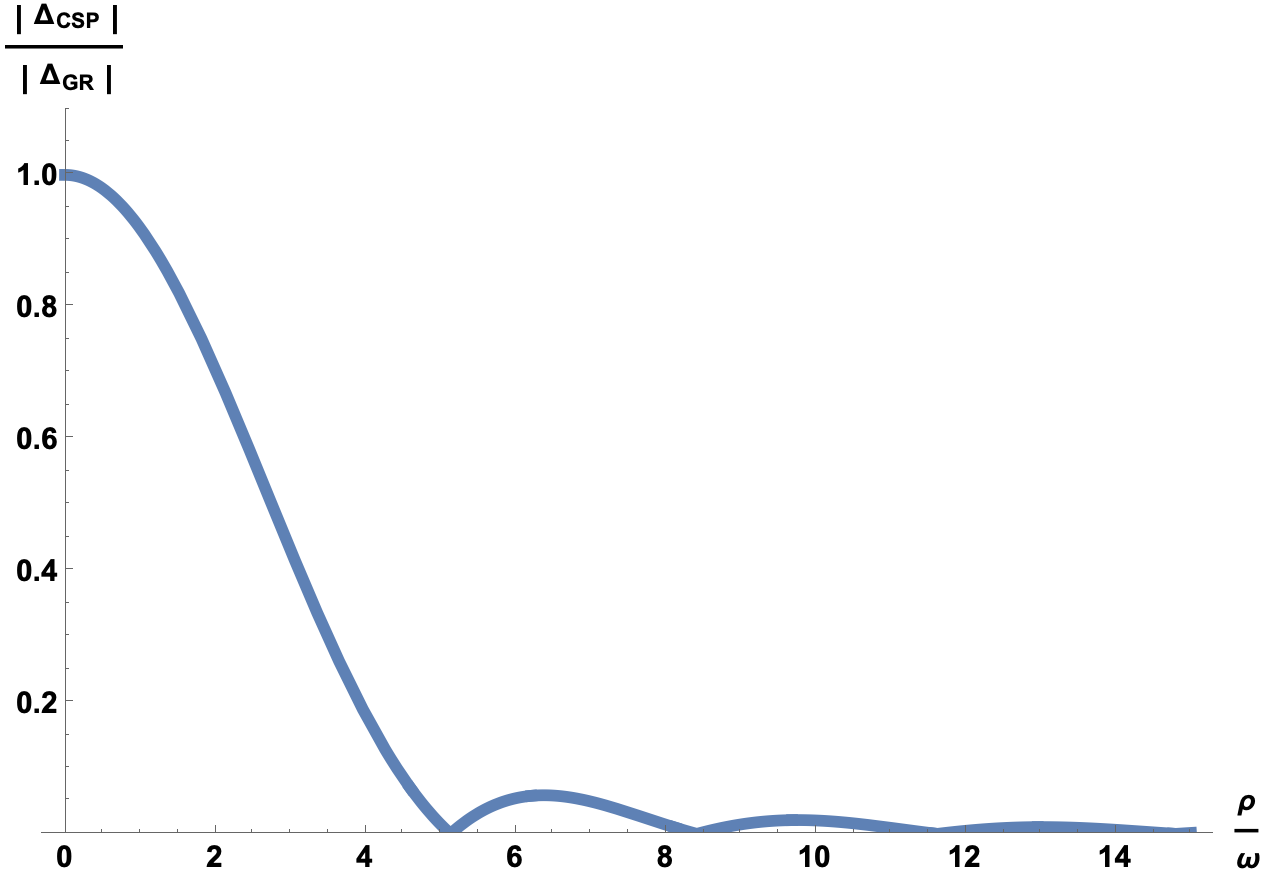}
\caption{The ratio of the amplitude of the strain for the CSP Gravity case ($\rh\neq 0$) to the GR strain as a function of $\rh/\om$. At $\om \gg \rh$, we see that the strain approaches the usual strain in GR. At $\om \ll \rh$, the strain is suppressed and vanishes in the $\omega\rightarrow 0$ (or $\rho\rightarrow \infty$) limit. The intercepts with zero are physical: they correspond to values of $\rho/\omega$ for which the detector is completely unaffected by helicity-2 waves transverse to it. Note, however, that these ``sensitivity zeros'' will occur at different values of $\omega$ for waves impinging on the detector obliquely (in general, the argument of $J_2$ in \eqref{cspeome} depends on the angle of $\dot \gamma^{(0)}$ relative to $k$, and \eqref{simplifiedEinbeinRho} assumes a right angle).}
\label{plot}
\end{center}
\end{figure} 

Consider the following ratio
\begin{equation}
    \frac{\Delta_{\text{CSP}}}{\Delta_{\text{GR}}} = 8 \left( \frac{\om}{\rh} \right)^2 J_2 \left( \frac{\rh}{\om} \right),
    \label{eq:CSP_suppression}
\end{equation}
where we have used \eqref{gwstrain}. We can see (see figure \ref{plot}) that the observed strain relative to GR is suppressed as the graviton spin-scale is increased, and in the $\rho\rightarrow \infty$ limit, the signal vanishes. Qualitatively, the reduction in the signal follows from the fact that the helicity states of a CSP are not boost invariant. Therefore, the relativistic photon ``sees" a mixture of helicity states, suppressing the observed signal from the quadrupole mode.

We highlight two observable consequences of this suppression, each of which can be dramatic depending on the scale of $\rho$:
The first is the dramatic suppression of signals from sources at frequencies $\omega\ll \rho$. For example, at $\rho=10^{-12}$ eV, the signal from inspirals at 100 Hz ($\rho/\omega \approx 3$) would be reduced by a factor of $\sim 2$ and therefore the volume in which such signals are detectable, and the expected number of observed sources, would fall by a factor of 8. For 35 Hz signals ($\rho/\omega \approx 10$), the number of detectable sources would fall by a factor of 1000!

The second effect, present even at $\rho/\omega \lesssim 1$, is the \emph{non-standard evolution} of the GW wave-form over time. As a binary system inspirals, its frequency increases --- the inspiral phase of well measured sources such as GW230814 have been measured over a sizable frequency range, from 30 to $\sim$ 100 Hz \cite{LIGOScientific:2025cmm}.  The amplitude of the resulting gravitational waves increases with frequency approximately as $\omega^{2/3}$.  If the graviton spin-scale $\rho$ is non-zero, then, in addition, the detector response to a given GW amplitude \emph{also} increases with $\omega$, as seen in \eqref{cspstrainexp} and Fig.~\ref{plot}, where increasing $\omega$ corresponds to moving leftward on the plot. The resulting change to the observed waveforms of binary mergers is fully predicted by \eqref{cspstrainexp} in terms of the single parameter $\rho$, and can be used to constrain smaller $\rho$ than those discussed above --- although care must be taken to avoid confusion with variation of the source parameters, and we will not undertake a detailed analysis of how large $\rho$ must be for this effect to be observably large.


\section{Future Directions}\label{4}
In this paper, we computed the gravitational time-delay for an idealized photon interferometer setup for the case where the graviton is a CSP particle. We demonstrated that for the $\e$-space interaction currents used in our analysis, the appropriate GR results are recovered in the $\rh \rightarrow 0$ limit. At non-zero $\rh$, the time-delay signature differs significantly only for frequencies $\omega\lesssim \rh$, and in particular is suppressed relative to the predictions of GR. These results represent the first predictions computed for gravity when the graviton spin-scale $\rh$ is non-zero, and we expect these results to be useful for measuring and/or constraining the graviton spin scale $\rh$ in ongoing and future GW experiments.  

For this calculation, we used two simplifications. The first is that we treat the photon as a scalar. While this is sufficient for our analysis here, the CSP coupling to photon (or matter) spin degrees of freedom would be interesting to investigate. Secondly, we used a monochromatic background wave composed only of $h=\pm 2$ modes, which ignores O($\rh$) corrections to the originating GW production mechanism. In principle, the underlying production process could create partner $h\neq \pm 2$ modes of the graviton at $O( \rh/\om )$, which is also the leading order at which the strain is affected in this calculation (see \eqref{cspstrainexp}). Computing and understanding the production of these partner modes in astrophysical sources is both tractable with the present formalism and would be very interesting to study. To compute a realistic waveform in sources, including non-linear effects, a more complete theory of interacting CSP's is required, which is presently a pressing direction of ongoing work. 

Even with the present limitations to the linear regime, it would be interesting to compute waveforms for binary mergers, and use them to predict signals that can be constrained. Since the CSP effect is parametrically controlled by $\rh/\om$, each detector is naturally sensitive to values of $\rh \sim \mathcal{O} \left( \om_{\text{s}} \right)$, where $\om_{\text{s}}$ is the characteristic frequency that the sources that the detector is sensitive to. Naturally, the lower the energy that a detector can probe, the stronger of a constraint can be put on the spin-scale for CSP gravity. The current set of ground based detectors, such as LIGO \cite{LIGOScientific:2016aoc}, VIRGO \cite{Virgo}, and KAGRA \cite{Kagra}, are sensitive to stellar mass black hole mergers at $\omega$ of roughly $10^{-15} - 10^{-13}$ eV ($10 - 10^3$ Hz). Therefore they can constrain $\rho$ in a similar range. Future ground based detectors such as Einstein Telescope \cite{Sathyaprakash:2009xt, Punturo:2010zz, Sathyaprakash:2012jk}, and Cosmic Explorer \cite{Reitze:2019iox, Evans:2021gyd, Mpetha:2022xqo} will primarily improve sensitivity to higher-frequency signals, where CSP effects are weaker. Thus, if the current ground based detectors do not show a hint of the CSP nature of gravity, then these future detectors will not be any more sensitive. On the other hand, space based interferometers such as LISA \cite{LISACosmologyWorkingGroup:2022kbp, LISACosmologyWorkingGroup:2019mwx, LISA:2022kgy, LISA:2017pwj}, and TianQin \cite{TianQin} will probe much lower-frequency signals such as black hole binary mergers \cite{LISAConsortiumWaveformWorkingGroup:2023arg} at frequencies of roughly $10^{-20} - 10^{-16}$ eV ($10^{-4} - 1$ Hz), expanding sensitivity to comparably small $\rho$.

One can also look for CSP signatures in the stochastic gravitational wave background, which are measured using Pulsar Timing Array's (PTA's) \cite{Detweiler:1979wn, Sazhin:1978myk}. Recent data releases from PTA groups around the world \cite{NANOGrav:2023gor, NANOGrav:2023hde, EPTA:2023sfo, EPTA:2023gyr, EPTA:2023xxk, EPTA:2023fyk, EPTA:2023akd, Reardon:2023gzh, Reardon:2023zen, Zic:2023gta, Xu:2023wog} demonstrate an angular correlation that is suggestive of the Hellings-Downs predictions consistent with GR, but with significant error bars. Computing the size of non-quadrupole signatures due to a CSP gravitational wave background could potentially open another avenue for measuring or constraining the gravitational spin-scale $\rh$. 


\begin{acknowledgments}
We thank Shoaib Akhtar, Kevin Zhou, Gowri Sunderesan, and Aidan Reilly for general discussions and feedback. The authors are supported by the U.S. Department of Energy under contract number DE-AC02-76SF00515.
\end{acknowledgments}


\appendix


\section{GW Gauge Terms}\label{A}
\subsection{Gauge Terms in Linearized GR}
The Fierz-Pauli action, \eqref{action}, has the gauge symmetry (linearized diffeomorphism invariance) as noted in \eqref{fpgauge}. In Fourier space, the gauge symmetry implies that 
\begin{equation}
\label{fpkgauge}
    \ep_{\mn}' = \ep_{\mn} + k_\m \chi_\n + k_\n \chi_\m
\end{equation}
where $k_\m = (\om, 0, 0, \om)$ is the 4-vector for a wave propagating along the $z$-axis (as we fixed for our GW), and $\chi_\m$ is some arbitrary 4-vector. Combining the above expression with \eqref{ddgauge} gives us only two possible polarization states as defined in \eqref{pol}, up to some overall undetermined constant. For more detials on how this is done, the reader is referred to \cite{Feynman:1996kb}. The gauge terms are
\begin{equation}
\label{gwgaugeterms}
    h^{\text{gauge}}_{\mn} (z) = \sum_{i = 1,2,3} H^{(i)}_{\mn} h_i \cos (k \cdot z + \phi^{(i)})
\end{equation}
Where $\phi^{(i)}$, and $h_i$ are arbitrary phases and amplitudes, respectively, associated with each gauge term. For consistency, the gauge terms are monochromatic, with amplitudes comparable to $h_{+}$, and $h_{\cross}$. We also define $H^{(i)}_{\mn}$, which are "polarization" tensors for the gauge terms. 
\begin{align}
\label{gaugepol}
    H^{(1)} = \begin{pmatrix}
        1 & 0 & 0 & 0 \\
        0 & 0 & 0 & 0 \\
        0 & 0 & 0 & 0 \\
        0 & 0 & 0 & -1 \\
    \end{pmatrix} &\quad 
    H^{(2)} = \frac{1}{2} \begin{pmatrix}
        0 & -1 & 0 & 0 \\
        -1 & 0 & 0 & 1 \\
        0 & 0 & 0 & 0 \\
        0 & 1 & 0 & 0 \\
    \end{pmatrix} \quad 
    \nonumber \\
    H^{(3)} = &\frac{1}{2} \begin{pmatrix}
        0 & 0 & -1 & 0 \\
        0 & 0 & 0 & 0 \\
        -1 & 0 & 0 & 1 \\
        0 & 0 & 1 & 0 \\
    \end{pmatrix} \quad 
\end{align}

\subsection{Equations of Motion and Boundary Conditions with Gauge Terms}
When considering the gauge terms \eqref{gwgaugeterms}, and \eqref{gaugepol} along with the usual polarization states \eqref{pol}, the equation of motion, einbein constraint, and the geodesics have the same form as in \eqref{gweom}, \eqref{gweome} and \eqref{gwgeodesics} respectively, with the sum also including the gauge terms.

The photon geodesic in the presence of gauge terms has a total of $ 2 (5+1) = 12$ undetermined 4-vector coefficients (here, following the parameter-counting below \eqref{gwgeodesics}, we have $a=5$ accounting for terms linear in the 3 gauge ``polarizations'' as well as the 2 physical ones). These are fixed by matching with the mirrors as was done in \ref{2.3} - term by term for each perturbation parameter. 

In section \ref{2.3}, the mirror geodesics were fixed by leveraging the fact that we were free to choose the detector rest frame. Including the gauge terms is a departure from the detector rest frame, as we pick up sinusoidal terms proportional to the gauge terms in the geodesics. This motivates us to define the boundary condition as a time average.
\begin{equation}
    \langle \mathcal{M}_{1i} \rangle = \lim_{\tau' \rightarrow \infty}\frac{1}{\tau'} \int_0^{\tau'} \dd{\tau_1} \mathcal{M}_{1i} (\tau_1) = L_{1i}.
\end{equation}
Here $\mathcal{M}_{1i} (\tau_1 = 0) = L_{1i}$ is the initial position of the mirror before the passage of the gravitational wave. This averaging kills the sinusoidal terms in the geodesics, but keeps the gauge terms. This tells us that the seemingly ad-hoc choice that we made to choose the detector rest frame is justified. The survival of the gauge terms should not be surprising -- the observed geodesics are frame dependent. 

The time-delay, however, is not dependent on the gauge terms. While the einbein constraint \eqref{gweome} picks up gauge terms, when the full calculation is done, to leading order in the perturbative expansion, the time-delay is independent of the gauge terms. This serves as a check on the calculation. This analysis can be extended to the CSP case, where the gauge terms can be taken from equation 2.24 of \cite{Schuster:2023xqa}, and in that case too, the answer is independent of the gauge terms.


\section{Photon Geodesics}\label{B}
\subsection{GW Background}
Here we list the photon geodesics for the GW case. These are the photons corresponding to $\g_e (\tau_e)$, and $\g_r (\tau_r)$ as shown in \ref{setup}, with boundary conditions, and the einbein constraint imposed.  We start with the emitted photon, with the time geodesic $ t_{(e)} (\tau_e)$, and the 3-vector of spatial geodesics $\vec{\g}_{(e)} (\tau_e)$. 
\begin{align}
\label{ge1}
    t_{(e)} (\tau_e) &= L \left( \frac{\tau_e}{\tau_{*}} \right) + h_{+} \left( g_{e} (0, \tau_e) + \frac{ \sin \left(L \omega \left( \frac{\tau_e}{\tau_{*}} \right) \right)}{ 2 \omega } \right)\\
\label{ge2}
    \vec{\g}_{(e)} (\tau_e) &= L \left( \frac{\tau_e}{\tau_{*}} \right) (1, 0, 0) + h_{+} g_{e} (0, \tau_e) \left( 1, 0, \frac{1}{2} \right)  \\
    \text{where -- } &g_{e} (0, \tau_e) = \frac{ \sin (L \omega) }{\omega} \left( \frac{\tau_e}{\tau_{*}} \right) - \frac{ \sin \left(L \omega \left( \frac{\tau_e}{\tau_{*}} \right) \right)}{ \omega }  \nonumber
\end{align}
Since $g_{e} (0,0) = g_{e} (0,\tau_{*}) = 0$, and $\vec{\g}_{(e)} (0) = (0, 0, 0)$, and $\tau = \tau_*$, $\vec{\g}_{(e)} (\tau_{*}) = (L, 0, 0)$, which matches the spatial coordinates of $\mathcal{M}_1$, and $\mathcal{M}_2$ respectively. As claimed earlier, the value of $\tau_{*}$ does not matter -- it scales how the affine parameter for the photon, $\tau_e$, behaves.  

Next, we consider the photons on the return trip, $\g_{(r)} (\tau_r)$
\begin{align}
\label{gr1}
    t_{(r)} (\tau_r) &= L \left( 1 + \frac{\tau_r}{\tau_\bullet} \right) \nonumber \\
    & \quad + h_{+} \left( g_{r} (\tau_r) + \frac{ \sin \left(L \omega \left( 1 + \frac{\tau_r}{\tau_f} \right) \right)}{2 \omega }\right) \\
\label{gr2}
    \vec{\g}_{(r)} (\tau_r) &= L \left( 1 - \frac{\tau_r}{\tau_\bullet} \right) (1, 0, 0) + h_{+} g_{r} (\tau_r) \left( 1, 0, \frac{1}{2} \right) \\
    \text{where} \rightarrow & g_{r} (0,\tau_r) = \frac{\sin (L \omega) }{\omega} - \frac{ \sin \left(L \omega \left( 1 + \frac{\tau_r}{\tau_\bullet} \right) \right)}{\omega } \nonumber \\
    & \qquad \qquad - \left( \frac{\tau_r}{\tau_\bullet} \right) \left( \frac{\sin (L \omega) }{\omega} - \frac{\sin (2L \omega) }{\omega}\right) \nonumber
\end{align}  
Again, $g_{r} (0,0) = g_{r} (0, \tau_\bullet) = 0$, so the spatial geodesics match the respective mirror geodesics. The time-delay is proportional to the $h_+$ term that survives at $\tau_r = \tau_\bullet$, and this matches the value calculated using the einbein constraint in \eqref{gwstrain}.

\subsection{CSP GW Background}
Here, we list the $t$ and $x$ geodesics for a CSP background. For the emitted photon, we have,
\begin{align}
    &t_{(e)} (\tau_e) = L \left( \frac{ \tau_e }{\tau_{*}} \right) + 2 h_{+} g_{e} (\rho, \tau_e) \nonumber \\
    &\qquad \qquad  + 4 h_{+} \left( \frac{\omega}{\rho^2} \right) J_2\left(\frac{\rho }{\omega }\right) \sin \left( L \omega \left( \frac{ \tau_e }{\tau_{*}}\right) \right) \\
    &x_{(e)} (\tau_e) = L \left( \frac{ \tau_e }{\tau_{*}} \right) + 2 h_{+} g_{e} (\rho, \tau_e) \\
    \text{where} \rightarrow & g_{e} (\rho, \tau_e) = g_{e} (0, \tau_e) \Bigg( 2 J_2 \left( \frac{\rho}{\omega} \right) \nonumber \\
     & \qquad \qquad \qquad + \left( \frac{\omega}{\rho} \right) \left(  J_1 \left(\frac{\rho }{\omega }\right) - J_3 \left(\frac{\rho }{\omega }\right) \right) \Bigg) \nonumber
\end{align} 
On taking the $\rh \rightarrow 0$ limit, these recover \eqref{ge1}, and \eqref{ge2}. Given that that the $\tau_{e}$ behavior of $g_{e} (\rho, \tau_e)$ is essentially $g_{e} (0, \tau_e)$, the $x$ geodesic of the photons agrees with the respective mirror.

For the return trip, the geodesics are $t_{(r)} (\tau_r)$, and $x_{(r)} (\tau_r)$.
\begin{align}
    &t_{(r)} (\tau_r) = L \left( 1 + \frac{\tau_r}{\tau_\bullet} \right) + 2 h_{+} g_{r} (\rho, \tau_r) \nonumber \\
    & \qquad \qquad 4 h_{+} \left( \frac{\omega}{\rho^2} \right) J_2 \left(\frac{\rho }{\omega }\right) \sin \left(L \omega \left( 1 + \frac{\tau_r}{\tau_\bullet} \right) \right) \\
    &x_{(r)} (\tau_r) = L \left( 1 - \frac{\tau_r}{\tau_\bullet} \right) + 2 h_{+} g_{r} (\rho, \tau_r)  \\
    \text{where} \rightarrow & g_{r} (\rho, \tau_r) = g_{r} (0, \tau_r) \Bigg( 2 J_2 \left( \frac{\rho}{\omega} \right) \nonumber \\
     & \qquad \qquad \qquad + \left( \frac{\omega}{\rho} \right) \left(  J_1 \left(\frac{\rho }{\omega }\right) - J_3 \left(\frac{\rho }{\omega }\right) \right) \Bigg) \nonumber
\end{align}
Again, in the $\rho \rightarrow 0$ limit, these recover \eqref{gr1}, and \eqref{gr2}. It is easy to check that this matches the respective mirror geodesics. The time-delay can be read off from the term proportional to $h_+$ that is non-vanishing at $\tau_r = \tau_\bullet$, and this agrees with \eqref{cspstrain}.


\section{CSP - Identities and EoM}\label{C}
\subsection{4-Vector Contraction Identities} \label{C.1}
Consider the CSP interaction term, 
\begin{align}
\label{appendixint}
    S_{\text{int}} &= h_{+} \int \frac{\dd{\tau}}{e} \left( \gd^2 - 2 \left( \frac{\kgd}{\rh} \right)^{2} \right) J_2 \left( \sqrt{-V^{2}}\right) \cos (\kg) \nonumber \\
    & \qquad \qquad \times \left( \left( \frac{\ep^{+} \cdot V}{|\ep^{+} \cdot V|} \right)^2 + \left( \frac{\ep^{-} \cdot V}{|\ep^{-} \cdot V|} \right)^2 \right) 
\end{align}
The contraction of $V^\m$ (defined in \eqref{V}) with $\ep^{\pm}_\m$ are
\begin{equation}
    \ep^{\pm} \cdot V = - \rh \frac{\ep^{\pm} \cdot \gd}{\kgd} \text{, } \ep^{\pm} \cdot V = \left( \frac{\rh}{\kgd} \right) \sqrt{(\ep^{-} \cdot \gd) (\ep^{+} \cdot \gd)}
\end{equation}
This simplifies the second term in \eqref{appendixint}.
\begin{equation}
    \left( \frac{\ep^{+} \cdot V}{|\ep^{+} \cdot V|} \right)^2 + \left( \frac{\ep^{-} \cdot V}{|\ep^{-} \cdot V|} \right)^2 = \frac{\ep^{+} \cdot \gd}{\ep^{-} \cdot \gd} + \frac{\ep^{-} \cdot \gd}{\ep^{+} \cdot \gd} 
\end{equation}
Lastly, recalling that the $\mathbf{g}_{\mn} = \left( q_\m k_\n + q_\n k_\m \right) - (1/2) \left( \ep_{+}^\m \ep_{-}^\n + \ep_{+}^\n \ep_{-}^\m \right)$, allows us to express $\gd^2 = 2 \left( \qgd \right) \left( \kgd \right) - \left( \ep^{(+)} \cdot \gd \right) \left( \ep^{(-)} \cdot \gd \right)$. The argument of the Bessel function is
\begin{align}
   \sqrt{ - V^2 } &= \sqrt{- \left( \frac{\rh}{\kgd} \right)^2 \bigg( \gd^2 - 2 (\qgd) (\kgd) \bigg)} \nonumber \\
   &= \left( \frac{\rh}{\kgd} \right) \sqrt{(\ep^{+} \cdot \gd ) (\ep^{-} \cdot \gd) }
\end{align}
This simplifies the integrand of the interaction term
\begin{align}
    &\frac{h_{+}}{e} \left( \gd^2 - 2 \left( \frac{\kgd}{\rh} \right)^{2} \right) \left( \frac{\ep^{+} \cdot \gd}{\ep^{-} \cdot \gd} + \frac{\ep^{-} \cdot \gd}{\ep^{+} \cdot \gd}  \right) \nonumber \\
    & \quad \times J_2 \left( \left( \frac{\rh}{\kgd} \right) \sqrt{(\ep^{+} \cdot \gd ) (\ep^{-} \cdot \gd)} \right) \cos (\kg)
\end{align}

\subsection{CSP EoM} \label{C.2}
The CSP EoM can be found in \eqref{cspeom}. Recalling the definitions of $\z = \ep^{+} \cdot \gd$, and $\zb = \ep^{-} \cdot \gd$, we further define $\z^{(0)} = \ep^{+} \cdot \gd^{(0)}$, and $\zb^{(0)} = \ep^{-} \cdot \gd^{(0)}$. Using this, we express the $\rh$ deformed $A^{(+)}_\m$ function in \eqref{cspeom}.
\begin{widetext}
\begin{align}
    &2 h_{+} \sin \left( \kg^{(0)} \right) \left[ - \left( \kgd^{(0)} \right) \left\{ \left( \gd^{(0)}_\m - \frac{2 \left( \kgd^{(0)} \right) k_\m}{ \rh^2} \right) \left( \frac{\z^{(0)}}{\zb^{(0)}} + \frac{\zb^{(0)}}{\z^{(0)}}  \right) \right. \right. \nonumber \\
    &\left. \left. + \left( \frac{ \left(\gd^{(0)} \right)^2 }{2} - \frac{\left( \kgd^{(0)} \right)^2}{ \rh^2} \right) \left( \frac{\ep^{+}_\m}{\zb^{(0)} } - \frac{\z^{(0)} \ep^{-}_\m }{\left( \zb^{(0)} \right)^2} + \frac{\ep^{-}_\m}{ \z^{(0)} } - \frac{ \zb^{(0)} \ep^{+}_\m }{\left( \z^{(0)} \right)^2} \right) \right. \right. \nonumber \\
    &\left. \left. + \frac{k_\m}{ \left( \kgd^{(0)} \right)} \left( \frac{ \left(\gd^{(0)} \right)^2 }{2} - \frac{\left( \kgd^{(0)} \right)^2}{ \rh^2} \right) \left( \frac{\z^{(0)}}{\zb^{(0)}} + \frac{\zb^{(0)}}{\z^{(0)}}  \right) \right\} J_2 \left( \frac{\rh}{\kgd^{(0)}} \sqrt{ \z^{(0)} \zb^{(0)} } \right) \right. \nonumber \\
    & \left. - \frac{1}{2} \left( \frac{ \left(\gd^{(0)} \right)^2 }{2} - \frac{\left( \kgd^{(0)} \right)^2}{ \rh^2} \right) \left( \frac{\ep^{+} \cdot \gd^{(0)}}{\ep^{-} \cdot \gd^{(0)}} + \frac{\ep^{-} \cdot \gd^{(0)}}{\ep^{+} \cdot \gd^{(0)}}  \right) \left( J_1 \left( \frac{\rh}{\kgd^{(0)}} \sqrt{ \z^{(0)} \zb^{(0)} } \right)\right. \right. \nonumber \\
    & \left. \left. - J_3 \left( \frac{\rh}{\kgd^{(0)}} \sqrt{ \z^{(0)} \zb^{(0)} } \right) \right) \left( - \frac{\rh k_\m}{\kgd^{(0)}} \sqrt{ \z^{(0)} \zb^{(0)} } + \frac{1}{2} \sqrt{ \frac{\z^{(0)}}{\zb^{(0)}} } \ep^-_\m + \frac{1}{2} \sqrt{ \frac{\zb^{(0)}}{\z^{(0)}} } \ep^+_\m \right)  \right]
\end{align}
\end{widetext}
For massive particles, the argument of the Bessel functions are vanishing since $\z^{(0)} = \zb^{(0)} = 0$, and this sets $A^{(+)}_\m = 0$. 

For photons, there are simplifications for our set up. Since the photons are constrained to move along the $x$-axis, $\z^{(0)} = \zb^{(0)}$. The orthogonality of the photon to the background GW gives a constraint -- $\vec{k} \cdot \vec{\gd} = 0$. From this, we get that $\kgd^{(0)} = \om \sqrt{ \z^{(0)} \zb^{(0)} }$. Lastly, photons are massless, so $\left( \gd^{(0)} \right)^2 = 0$. This gives us the $\rh$ deformed $A^{(+)}_\m$ for photons.
\begin{widetext}
\begin{align}
    2 h_{+} \sin \left( \kg^{(0)} \right) &\left[ - 2 \left( \kgd^{(0)} \right) \left\{ - \frac{k_\m}{\rh^2} \left( \kgd^{(0)} \right)  + \left( \gd^{(0)}_\m - \frac{2 \left( \kgd^{(0)} \right) k_\m}{ \rh^2} \right) \right\} J_2 \left( \frac{\rh}{\om} \right) \right. \nonumber \\
    & \qquad \qquad  \left. + \left( \frac{\kgd^{(0)}}{\rh} \right)^2 \left( - \frac{\rh}{\om} k_\m + \frac{\ep^-_\m}{2} + \frac{\ep^+_\m}{2} \right) \left( J_1 \left( \frac{\rh}{\om} \right) - J_3 \left( \frac{\rh}{\om} \right) \right) \right]
\end{align}
\end{widetext}


\nocite{*}

\bibliography{ref}

@PREAMBLE{
 "\providecommand{\noopsort}[1]{}" 
 # "\providecommand{\singleletter}[1]{#1}%" 
}

@article{Buchbinder:2022msd,
    author = "Buchbinder, I. L. and Fedoruk, S. A. and Isaev, A. P. and Krykhtin, V. A.",
    title = "{On the off-shell superfield Lagrangian formulation of 4D, N=1 supersymmetric infinite spin theory}",
    eprint = "2203.12904",
    archivePrefix = "arXiv",
    primaryClass = "hep-th",
    doi = "10.1016/j.physletb.2022.137139",
    journal = "Phys. Lett. B",
    volume = "829",
    pages = "137139",
    year = "2022"
}

@article{Najafizadeh:2021dsm,
    author = "Najafizadeh, Mojtaba",
    title = "{Off-shell supersymmetric continuous spin gauge theory}",
    eprint = "2112.10178",
    archivePrefix = "arXiv",
    primaryClass = "hep-th",
    reportNumber = "IPM/P-2021/42",
    doi = "10.1007/JHEP02(2022)038",
    journal = "JHEP",
    volume = "02",
    pages = "038",
    year = "2022"
}

@article{Metsaev:2021zdg,
    author = "Metsaev, R. R.",
    title = "{Mixed-symmetry continuous-spin fields in flat and AdS spaces}",
    eprint = "2105.11281",
    archivePrefix = "arXiv",
    primaryClass = "hep-th",
    reportNumber = "FIAN-TD-2021-05",
    doi = "10.1016/j.physletb.2021.136497",
    journal = "Phys. Lett. B",
    volume = "820",
    pages = "136497",
    year = "2021"
}

@article{Buchbinder:2020nxn,
    author = "Buchbinder, I. L. and Fedoruk, S. and Isaev, A. P. and Krykhtin, V. A.",
    title = "{Towards Lagrangian construction for infinite half-integer spin field}",
    eprint = "2005.07085",
    archivePrefix = "arXiv",
    primaryClass = "hep-th",
    doi = "10.1016/j.nuclphysb.2020.115114",
    journal = "Nucl. Phys. B",
    volume = "958",
    pages = "115114",
    year = "2020"
}

@article{Najafizadeh:2019mun,
    author = "Najafizadeh, Mojtaba",
    title = "{Supersymmetric Continuous Spin Gauge Theory}",
    eprint = "1912.12310",
    archivePrefix = "arXiv",
    primaryClass = "hep-th",
    reportNumber = "IPM/P-2019/046",
    doi = "10.1007/JHEP03(2020)027",
    journal = "JHEP",
    volume = "03",
    pages = "027",
    year = "2020"
}

@article{Buchbinder:2019sie,
    author = "Buchbinder, I. L. and Isaev, A. P. and Fedoruk, S. A.",
    title = "{Massless Infinite Spin (Super)particles and Fields}",
    eprint = "1911.00362",
    archivePrefix = "arXiv",
    primaryClass = "hep-th",
    doi = "10.1134/S0081543820030049",
    journal = "Proc. Steklov Inst. Math.",
    volume = "309",
    number = "1",
    pages = "46--56",
    year = "2020"
}

@article{Burdik:2019tzg,
    author = "Burd\'\i{}k, \v{C}. and Pandey, V. K. and Reshetnyak, A.",
    title = "{BRST\textendash{}BFV and BRST\textendash{}BV descriptions for bosonic fields with continuous spin on $R^{1,d-1}$}",
    eprint = "1906.02585",
    archivePrefix = "arXiv",
    primaryClass = "hep-th",
    doi = "10.1142/S0217751X20501547",
    journal = "Int. J. Mod. Phys. A",
    volume = "35",
    number = "26",
    pages = "2050154",
    year = "2020"
}

@article{Buchbinder:2019kuh,
    author = "Buchbinder, I. L. and Khabarov, M. V. and Snegirev, T. V. and Zinoviev, Yu. M.",
    title = "{Lagrangian formulation for the infinite spin $N$=1 supermultiplets in $d$=4}",
    eprint = "1904.05580",
    archivePrefix = "arXiv",
    primaryClass = "hep-th",
    doi = "10.1016/j.nuclphysb.2019.114717",
    journal = "Nucl. Phys. B",
    volume = "946",
    pages = "114717",
    year = "2019"
}

@article{Metsaev:2019opn,
    author = "Metsaev, R. R.",
    title = "{Light-cone continuous-spin field in AdS space}",
    eprint = "1903.10495",
    archivePrefix = "arXiv",
    primaryClass = "hep-th",
    reportNumber = "FIAN-TD-2019-04",
    doi = "10.1016/j.physletb.2019.04.041",
    journal = "Phys. Lett. B",
    volume = "793",
    pages = "134--140",
    year = "2019"
}

@article{Buchbinder:2019esz,
    author = "Buchbinder, I. L. and Gates, S. James and Koutrolikos, K.",
    title = "{Superfield continuous spin equations of motion}",
    eprint = "1903.08631",
    archivePrefix = "arXiv",
    primaryClass = "hep-th",
    doi = "10.1016/j.physletb.2019.05.015",
    journal = "Phys. Lett. B",
    volume = "793",
    pages = "445--450",
    year = "2019"
}

@article{Metsaev:2018moa,
    author = "Metsaev, R. R.",
    title = "{Cubic interaction vertices for massive/massless continuous-spin fields and arbitrary spin fields}",
    eprint = "1809.09075",
    archivePrefix = "arXiv",
    primaryClass = "hep-th",
    reportNumber = "FIAN-TD-2018-19",
    doi = "10.1007/JHEP12(2018)055",
    journal = "JHEP",
    volume = "12",
    pages = "055",
    year = "2018"
}

@article{Alkalaev:2018bqe,
    author = "Alkalaev, Konstantin and Chekmenev, Alexander and Grigoriev, Maxim",
    title = "{Unified formulation for helicity and continuous spin fermionic fields}",
    eprint = "1808.09385",
    archivePrefix = "arXiv",
    primaryClass = "hep-th",
    reportNumber = "FIAN-TD-2018-14",
    doi = "10.1007/JHEP11(2018)050",
    journal = "JHEP",
    volume = "11",
    pages = "050",
    year = "2018"
}

@article{Rivelles:2018tpt,
    author = "Rivelles, Victor O.",
    title = "{A Gauge Field Theory for Continuous Spin Tachyons}",
    journal = {arXiv e-prints},
    eprint = "1807.01812",
    archivePrefix = "arXiv",
    primaryClass = "hep-th",
    month = "7",
    year = "2018"
}

@article{Buchbinder:2018yoo,
    author = "Buchbinder, I. L. and Krykhtin, V. A. and Takata, H.",
    title = "{BRST approach to Lagrangian construction for bosonic continuous spin field}",
    eprint = "1806.01640",
    archivePrefix = "arXiv",
    primaryClass = "hep-th",
    doi = "10.1016/j.physletb.2018.07.070",
    journal = "Phys. Lett. B",
    volume = "785",
    pages = "315--319",
    year = "2018"
}

@article{Metsaev:2018lth,
    author = "Metsaev, R. R.",
    title = "{BRST-BV approach to continuous-spin field}",
    eprint = "1803.08421",
    archivePrefix = "arXiv",
    primaryClass = "hep-th",
    reportNumber = "FIAN-TD-2018-03",
    doi = "10.1016/j.physletb.2018.04.038",
    journal = "Phys. Lett. B",
    volume = "781",
    pages = "568--573",
    year = "2018"
}

@article{Alkalaev:2017hvj,
    author = "Alkalaev, Konstantin B. and Grigoriev, Maxim A.",
    title = "{Continuous spin fields of mixed-symmetry type}",
    eprint = "1712.02317",
    archivePrefix = "arXiv",
    primaryClass = "hep-th",
    reportNumber = "FIAN-TD-2017-28",
    doi = "10.1007/JHEP03(2018)030",
    journal = "J. Phys. A",
    volume = "03",
    pages = "030",
    year = "2018"
}

@article{LIGOScientific:2025cmm,
    author = "LIGO Scientific Collaboration and VIRGO and Kagra",
    title = "{GW230814: investigation of a loud gravitational-wave signal observed with a single detector}",
    eprint = "12509.07348",
    archivePrefix = "arXiv",
    primaryClass = "gr-qc",
    reportNumber = "LIGO-P230814",
    journal = "",
    month = "9",
    year = "2025"
}

@article{Metsaev:2017myp,
    author = "Metsaev, R. R.",
    title = "{Continuous-spin mixed-symmetry fields in AdS(5)}",
    eprint = "1711.11007",
    archivePrefix = "arXiv",
    primaryClass = "hep-th",
    reportNumber = "FIAN-TD-2017-27",
    doi = "10.1088/1751-8121/aabcda",
    journal = "J. Phys. A",
    volume = "51",
    number = "21",
    pages = "215401",
    year = "2018"
}

@article{Khabarov:2017lth,
    author = "Khabarov, M. V. and Zinoviev, Yu. M.",
    title = "{Infinite (continuous) spin fields in the frame-like formalism}",
    eprint = "1711.08223",
    archivePrefix = "arXiv",
    primaryClass = "hep-th",
    doi = "10.1016/j.nuclphysb.2018.01.016",
    journal = "Nucl. Phys. B",
    volume = "928",
    pages = "182--216",
    year = "2018"
}

@article{Bekaert:2017xin,
    author = "Bekaert, Xavier and Mourad, Jihad and Najafizadeh, Mojtaba",
    title = "{Continuous-spin field propagator and interaction with matter}",
    eprint = "1710.05788",
    archivePrefix = "arXiv",
    primaryClass = "hep-th",
    reportNumber = "IPM/P-2017/094",
    doi = "10.1007/JHEP11(2017)113",
    journal = "JHEP",
    volume = "11",
    pages = "113",
    year = "2017"
}

@article{Metsaev:2017cuz,
    author = "Metsaev, R. R.",
    title = "{Cubic interaction vertices for continuous-spin fields and arbitrary spin massive fields}",
    eprint = "1709.08596",
    archivePrefix = "arXiv",
    primaryClass = "hep-th",
    reportNumber = "FIAN-TD-2017-19",
    doi = "10.1007/JHEP11(2017)197",
    journal = "JHEP",
    volume = "11",
    pages = "197",
    year = "2017"
}

@article{Bekaert:2017khg,
    author = "Bekaert, Xavier and Skvortsov, Evgeny D.",
    title = "{Elementary particles with continuous spin}",
    eprint = "1708.01030",
    archivePrefix = "arXiv",
    primaryClass = "hep-th",
    doi = "10.1142/S0217751X17300198",
    journal = "Int. J. Mod. Phys. A",
    volume = "32",
    number = "23n24",
    pages = "1730019",
    year = "2017"
}

@article{Najafizadeh:2017tin,
    author = "Najafizadeh, Mojtaba",
    title = "{Modified Wigner equations and continuous spin gauge field}",
    eprint = "1708.00827",
    archivePrefix = "arXiv",
    primaryClass = "hep-th",
    reportNumber = "IPM/P-2018/085",
    doi = "10.1103/PhysRevD.97.065009",
    journal = "Phys. Rev. D",
    volume = "97",
    number = "6",
    pages = "065009",
    year = "2018"
}

@article{Zinoviev:2017rnj,
    author = "Zinoviev, Yu. M.",
    title = "{Infinite spin fields in d = 3 and beyond}",
    eprint = "1707.08832",
    archivePrefix = "arXiv",
    primaryClass = "hep-th",
    doi = "10.3390/universe3030063",
    journal = "Universe",
    volume = "3",
    number = "3",
    pages = "63",
    year = "2017"
}

@article{Metsaev:2017ytk,
    author = "Metsaev, R. R.",
    title = "{Fermionic continuous spin gauge field in (A)dS space}",
    eprint = "1703.05780",
    archivePrefix = "arXiv",
    primaryClass = "hep-th",
    reportNumber = "FIAN-TD-2017-06",
    doi = "10.1016/j.physletb.2017.08.020",
    journal = "Phys. Lett. B",
    volume = "773",
    pages = "135--141",
    year = "2017"
}

@article{Metsaev:2016lhs,
    author = "Metsaev, R. R.",
    title = "{Continuous spin gauge field in (A)dS space}",
    eprint = "1610.00657",
    archivePrefix = "arXiv",
    primaryClass = "hep-th",
    reportNumber = "FIAN-TD-2016-22",
    doi = "10.1016/j.physletb.2017.02.027",
    journal = "Phys. Lett. B",
    volume = "767",
    pages = "458--464",
    year = "2017"
}

@article{Rivelles:2016rwo,
    author = "Rivelles, Victor O.",
    title = "{Remarks on a Gauge Theory for Continuous Spin Particles}",
    eprint = "1607.01316",
    archivePrefix = "arXiv",
    primaryClass = "hep-th",
    doi = "10.1140/epjc/s10052-017-4927-1",
    journal = "Eur. Phys. J. C",
    volume = "77",
    number = "7",
    pages = "433",
    year = "2017"
}

@article{Bekaert:2015qkt,
    author = "Bekaert, X. and Najafizadeh, M. and Setare, M. R.",
    title = "{A gauge field theory of fermionic Continuous-Spin Particles}",
    eprint = "1506.00973",
    archivePrefix = "arXiv",
    primaryClass = "hep-th",
    doi = "10.1016/j.physletb.2016.07.005",
    journal = "Phys. Lett. B",
    volume = "760",
    pages = "320--323",
    year = "2016"
}

@article{Rivelles:2014fsa,
    author = "Rivelles, Victor O.",
    title = "{Gauge Theory Formulations for Continuous and Higher Spin Fields}",
    eprint = "1408.3576",
    archivePrefix = "arXiv",
    primaryClass = "hep-th",
    doi = "10.1103/PhysRevD.91.125035",
    journal = "Phys. Rev. D",
    volume = "91",
    number = "12",
    pages = "125035",
    year = "2015"
}

@article{Schuster:2014xja,
    author = "Schuster, Philip and Toro, Natalia",
    title = "{A new class of particle in 2 + 1 dimensions}",
    eprint = "1404.1076",
    archivePrefix = "arXiv",
    primaryClass = "hep-th",
    doi = "10.1016/j.physletb.2015.02.050",
    journal = "Phys. Lett. B",
    volume = "743",
    pages = "224--227",
    year = "2015"
}

@article{Schuster:2014hca,
    author = "Schuster, Philip and Toro, Natalia",
    title = "{Continuous-spin particle field theory with helicity correspondence}",
    eprint = "1404.0675",
    archivePrefix = "arXiv",
    primaryClass = "hep-th",
    doi = "10.1103/PhysRevD.91.025023",
    journal = "Phys. Rev. D",
    volume = "91",
    pages = "025023",
    year = "2015"
}

@article{Font:2013hia,
    author = "Font, Anamaria and Quevedo, Fernando and Theisen, Stefan",
    title = "{A comment on continuous spin representations of the Poincare group and perturbative string theory}",
    eprint = "1302.4771",
    archivePrefix = "arXiv",
    primaryClass = "hep-th",
    reportNumber = "DAMTP-2013-10",
    doi = "10.1002/prop.201400067",
    journal = "Fortsch. Phys.",
    volume = "62",
    pages = "975--980",
    year = "2014"
}

@article{Schuster:2013pta,
    author = "Schuster, Philip and Toro, Natalia",
    title = "{A Gauge Field Theory of Continuous-Spin Particles}",
    eprint = "1302.3225",
    archivePrefix = "arXiv",
    primaryClass = "hep-th",
    reportNumber = "PI-PARTPHYS-317",
    doi = "10.1007/JHEP10(2013)061",
    journal = "JHEP",
    volume = "10",
    pages = "061",
    year = "2013"
}

@article{Schuster:2013vpr,
    author = "Schuster, Philip and Toro, Natalia",
    title = "{On the Theory of Continuous-Spin Particles: Helicity Correspondence in Radiation and Forces}",
    eprint = "1302.1577",
    archivePrefix = "arXiv",
    primaryClass = "hep-th",
    reportNumber = "PI-PARTPHYS-316",
    doi = "10.1007/JHEP09(2013)105",
    journal = "JHEP",
    volume = "09",
    pages = "105",
    year = "2013"
}

@article{Schuster:2013pxj,
    author = "Schuster, Philip and Toro, Natalia",
    title = "{On the Theory of Continuous-Spin Particles: Wavefunctions and Soft-Factor Scattering Amplitudes}",
    eprint = "1302.1198",
    archivePrefix = "arXiv",
    primaryClass = "hep-th",
    doi = "10.1007/JHEP09(2013)104",
    journal = "JHEP",
    volume = "09",
    pages = "104",
    year = "2013"
}

@article{Bekaert:2010hw,
    author = "Bekaert, Xavier and Boulanger, Nicolas and Sundell, Per",
    title = "{How higher-spin gravity surpasses the spin two barrier: no-go theorems versus yes-go examples}",
    eprint = "1007.0435",
    archivePrefix = "arXiv",
    primaryClass = "hep-th",
    doi = "10.1103/RevModPhys.84.987",
    journal = "Rev. Mod. Phys.",
    volume = "84",
    pages = "987--1009",
    year = "2012"
}

@article{Bekaert:2010hp,
    author = "Bekaert, Xavier and Boulanger, Nicolas and Leclercq, Serge",
    title = "{Strong obstruction of the Berends-Burgers-van Dam spin-3 vertex}",
    eprint = "1002.0289",
    archivePrefix = "arXiv",
    primaryClass = "hep-th",
    doi = "10.1088/1751-8113/43/18/185401",
    journal = "J. Phys. A",
    volume = "43",
    pages = "185401",
    year = "2010"
}

@article{Bekaert:2005in,
    author = "Bekaert, X. and Mourad, J.",
    title = "{The Continuous spin limit of higher spin field equations}",
    eprint = "hep-th/0509092",
    archivePrefix = "arXiv",
    reportNumber = "IHES-P-05-36",
    doi = "10.1088/1126-6708/2006/01/115",
    journal = "JHEP",
    volume = "01",
    pages = "115",
    year = "2006"
}

@article{Khan:2004nj,
    author = "Khan, Abu M. and Ramond, Pierre",
    title = "{Continuous spin representations from group contraction}",
    eprint = "hep-th/0410107",
    archivePrefix = "arXiv",
    reportNumber = "UFIFT-HEP-04-12",
    doi = "10.1063/1.1897663",
    journal = "J. Math. Phys.",
    volume = "46",
    pages = "053515",
    year = "2005",
    note = "[Erratum: J.Math.Phys. 46, 079901 (2005)]"
}

@article{Segal:2001di,
    author = "Segal, Arkady Yu.",
    title = "{Point particle in general background fields vsersus gauge theories of traceless symmetric tensors}",
    eprint = "hep-th/0110056",
    archivePrefix = "arXiv",
    reportNumber = "FIAN-TD-14-01",
    doi = "10.1142/S0217751X03015830",
    journal = "Int. J. Mod. Phys. A",
    volume = "18",
    pages = "4999--5021",
    year = "2003"
}

@article{Berends:1985xx,
    author = "Berends, Frits A. and Burgers, G. J. H. and van Dam, H.",
    title = "{Explicit construction of conserved currents for massless fields of arbitrary spin}",
    reportNumber = "Print-85-0830 (LEIDEN)",
    doi = "10.1016/S0550-3213(86)80019-0",
    journal = "Nucl. Phys. B",
    volume = "271",
    pages = "429--441",
    year = "1986"
}

@article{Berends:1984rq,
    author = "Berends, Frits A. and Burgers, G. J. H. and van Dam, H.",
    title = "{On the Theoretical Problems in Constructing Interactions Involving Higher Spin Massless Particles}",
    reportNumber = "IFP-234-UNC",
    doi = "10.1016/0550-3213(85)90074-4",
    journal = "Nucl. Phys. B",
    volume = "260",
    pages = "295--322",
    year = "1985"
}

@article{Weinberg:1980kq,
    author = "Weinberg, Steven and Witten, Edward",
    title = "{Limits on Massless Particles}",
    reportNumber = "HUTP-80/A056",
    doi = "10.1016/0370-2693(80)90212-9",
    journal = "Phys. Lett. B",
    volume = "96",
    pages = "59--62",
    year = "1980"
}

@article{deWit:1979sib,
    author = "de Wit, Bernard and Freedman, Daniel Z.",
    title = "{Systematics of Higher Spin Gauge Fields}",
    reportNumber = "Print-79-0659 (SUNY,STONY BROOK)",
    doi = "10.1103/PhysRevD.21.358",
    journal = "Phys. Rev. D",
    volume = "21",
    pages = "358",
    year = "1980"
}

@article{Fang:1978wz,
    author = "Fang, J. and Fronsdal, C.",
    title = "{Massless Fields with Half Integral Spin}",
    reportNumber = "UCLA/78/TEP/14",
    doi = "10.1103/PhysRevD.18.3630",
    journal = "Phys. Rev. D",
    volume = "18",
    pages = "3630",
    year = "1978"
}

@article{Fronsdal:1978rb,
    author = "Fronsdal, Christian",
    title = "{Massless Fields with Integer Spin}",
    reportNumber = "UCLA/78/TEP/5",
    doi = "10.1103/PhysRevD.18.3624",
    journal = "Phys. Rev. D",
    volume = "18",
    pages = "3624",
    year = "1978"
}

@article{Hirata:1977ss,
    author = "Hirata, K.",
    title = "{Quantization of Massless Fields with Continuous Spin}",
    doi = "10.1143/PTP.58.652",
    journal = "Prog. Theor. Phys.",
    volume = "58",
    pages = "652--666",
    year = "1977"
}

@article{Abbott:1976bb,
    author = "Abbott, L. F.",
    title = "{Massless Particles with Continuous Spin Indices}",
    reportNumber = "Print-76-0037 (BRANDEIS)",
    doi = "10.1103/PhysRevD.13.2291",
    journal = "Phys. Rev. D",
    volume = "13",
    pages = "2291",
    year = "1976"
}

@article{Singh:1974qz,
    author = "Singh, L. P. S. and Hagen, C. R.",
    title = "{Lagrangian formulation for arbitrary spin. 1. The boson case}",
    doi = "10.1103/PhysRevD.9.898",
    journal = "Phys. Rev. D",
    volume = "9",
    pages = "898--909",
    year = "1974"
}

@article{Singh:1974rc,
    author = "Singh, L. P. S. and Hagen, C. R.",
    title = "{Lagrangian formulation for arbitrary spin. 2. The fermion case}",
    doi = "10.1103/PhysRevD.9.910",
    journal = "Phys. Rev. D",
    volume = "9",
    pages = "910--920",
    year = "1974"
}

@article{Chakrabarti:1971rz,
    author = "Chakrabarti, A.",
    title = "{Remarks on lightlike continuous spin and spacelike representations of the poincare group}",
    doi = "10.1063/1.1665809",
    journal = "J. Math. Phys.",
    volume = "12",
    pages = "1813--1822",
    year = "1971"
}

@article{Iverson:1971hq,
    author = "Iverson, G. J. and Mack, G.",
    title = "{Quantum fields and interactions of massless particles - the continuous spin case}",
    doi = "10.1016/0003-4916(71)90284-3",
    journal = "Annals Phys.",
    volume = "64",
    pages = "211--253",
    year = "1971"
}

@article{Yngvason:1970fy,
    author = "Yngvason, J.",
    title = "{Zero-mass infinite spin representations of the poincare group and quantum field theory}",
    doi = "10.1007/BF01649432",
    journal = "Commun. Math. Phys.",
    volume = "18",
    pages = "195--203",
    year = "1970"
}

@article{Iverson:1970kyn,
    author = "Iverson, G. J. and Mack, G.",
    title = "{Theory of weak interactions with *continuous-spin* neutrinos}",
    doi = "10.1103/PhysRevD.2.2326",
    journal = "Phys. Rev. D",
    volume = "2",
    pages = "2326--2333",
    year = "1970"
}

@article{Weinberg:1965rz,
    author = "Weinberg, Steven",
    title = "{Photons and gravitons in perturbation theory: Derivation of Maxwell's and Einstein's equations}",
    doi = "10.1103/PhysRev.138.B988",
    journal = "Phys. Rev.",
    volume = "138",
    pages = "B988--B1002",
    year = "1965"
}

@article{Weinberg:1964ew,
    author = "Weinberg, Steven",
    title = "{Photons and Gravitons in  $S$-Matrix Theory: Derivation of Charge Conservation and Equality of Gravitational and Inertial Mass}",
    doi = "10.1103/PhysRev.135.B1049",
    journal = "Phys. Rev.",
    volume = "135",
    pages = "B1049--B1056",
    year = "1964"
}

@article{Weinberg:1964ev,
    author = "Weinberg, Steven",
    title = "{Feynman Rules for Any Spin. 2. Massless Particles}",
    doi = "10.1103/PhysRev.134.B882",
    journal = "Phys. Rev.",
    volume = "134",
    pages = "B882--B896",
    year = "1964"
}

@article{Wigner:1939cj,
    author = "Wigner, Eugene P.",
    editor = "Kim, Y. S. and Zachary, W. W.",
    title = "{On Unitary Representations of the Inhomogeneous Lorentz Group}",
    doi = "10.2307/1968551",
    journal = "Annals Math.",
    volume = "40",
    pages = "149--204",
    year = "1939"
}

@article{Schuster:2023xqa,
    author = "Schuster, Philip and Toro, Natalia and Zhou, Kevin",
    title = "{Interactions of Particles with ''Continuous Spin'' Fields}",
    eprint = "2303.04816",
    archivePrefix = "arXiv",
    primaryClass = "hep-th",
    reportNumber = "SLAC-PUB-17726",
    doi = "10.1007/JHEP04(2023)010",
    journal = "JHEP",
    volume = "04",
    pages = "010",
    year = "2023"
}

@article{Schuster:2023jgc,
    author = "Schuster, Philip and Toro, Natalia",
    title = "{Quantum electrodynamics mediated by a photon with continuous spin}",
    eprint = "2308.16218",
    archivePrefix = "arXiv",
    primaryClass = "hep-th",
    doi = "10.1103/PhysRevD.109.096008",
    journal = "Phys. Rev. D",
    volume = "109",
    number = "9",
    pages = "096008",
    year = "2024"
}

@article{Bellazzini:2024dco,
    author = "Bellazzini, Brando and De Angelis, Stefano and Romano, Marcello",
    title = "{Continuous-spin particles, on shell}",
    eprint = "2406.17017",
    archivePrefix = "arXiv",
    primaryClass = "hep-th",
    doi = "10.1007/JHEP05(2025)166",
    journal = "JHEP",
    volume = "05",
    pages = "166",
    year = "2025"
}

@article{Buchbinder:2024jpt,
    author = "Buchbinder, I. L. and Fedoruk, S. A. and Isaev, A. P. and Krykhtin, V. A.",
    title = "{BRST construction for infinite spin field on $AdS_4$}",
    eprint = "2403.14446",
    archivePrefix = "arXiv",
    primaryClass = "hep-th",
    doi = "10.1140/epjp/s13360-024-05430-6",
    journal = "Eur. Phys. J. Plus",
    volume = "139",
    number = "7",
    pages = "621",
    year = "2024"
}

@article{Buchbinder:2024hea,
    author = "Buchbinder, I. L. and Fedoruk, S. A. and Isaev, A. P. and Krykhtin, V. A.",
    title = "{Infinite (continuous) spin particle in constant curvature space}",
    eprint = "2402.13879",
    archivePrefix = "arXiv",
    primaryClass = "hep-th",
    doi = "10.1016/j.physletb.2024.138689",
    journal = "Phys. Lett. B",
    volume = "853",
    pages = "138689",
    year = "2024"
}

@book{Maggiore:2007ulw,
    author = "Maggiore, Michele",
    title = "{Gravitational Waves. Vol. 1: Theory and Experiments}",
    doi = "10.1093/acprof:oso/9780198570745.001.0001",
    isbn = "978-0-19-171766-6, 978-0-19-852074-0",
    publisher = "Oxford University Press",
    year = "2007"
}

@article{Grishchuk:1974jy,
    author = "Grishchuk, L. P.",
    title = "{Particle drift in the field of a gravitational wave}",
    journal = "Zh. Eksp. Teor. Fiz.",
    volume = "66",
    pages = "833--837",
    year = "1974"
}

@article{Detweiler:1979wn,
    author = "Detweiler, Steven L.",
    title = "{Pulsar timing measurements and the search for gravitational waves}",
    doi = "10.1086/157593",
    journal = "Astrophys. J.",
    volume = "234",
    pages = "1100--1104",
    year = "1979"
}

@article{Sazhin:1978myk,
    author = "Sazhin, Mikhail V.",
    title = "{Opportunities for detecting ultralong gravitational waves}",
    journal = "Sov. Astron.",
    volume = "22",
    pages = "36--38",
    year = "1978"
}

@article{Hellings:1983fr,
    author = "Hellings, R. w. and Downs, G. s.",
    title = "{UPPER LIMITS ON THE ISOTROPIC GRAVITATIONAL RADIATION BACKGROUND FROM PULSAR TIMING ANALYSIS}",
    doi = "10.1086/183954",
    journal = "Astrophys. J. Lett.",
    volume = "265",
    pages = "L39--L42",
    year = "1983"
}

@book{Misner:1973prb,
    author = "Misner, Charles W. and Thorne, K. S. and Wheeler, J. A.",
    title = "{Gravitation}",
    isbn = "978-0-7167-0344-0, 978-0-691-17779-3",
    publisher = "W. H. Freeman",
    address = "San Francisco",
    year = "1973"
}

@book{Carroll:2004st,
    author = "Carroll, Sean M.",
    title = "{Spacetime and Geometry}: {An Introduction to General Relativity}",
    doi = "10.1017/9781108770385",
    isbn = "978-0-8053-8732-2, 978-1-108-48839-6, 978-1-108-77555-7",
    publisher = "Cambridge University Press",
    month = "7",
    year = "2019"
}

@book{Feynman:1996kb,
    author = "Feynman, R. P.",
    editor = "Morinigo, F. B. and Wagner, W. G. and Hatfield, B.",
    title = "{Feynman lectures on gravitation}",
    publisher = "{CRC Press}",
    doi = "10.1201/9780429502859",
    isbn = "978-0-429-50285-9",
    year = "1996"
}

@article{Schuster:2024wjc,
    author = "Schuster, Philip and Sundaresan, Gowri and Toro, Natalia",
    title = "{Thermodynamics of continuous spin photons}",
    eprint = "2406.14616",
    archivePrefix = "arXiv",
    primaryClass = "hep-ph",
    doi = "10.1103/PhysRevD.111.056019",
    journal = "Phys. Rev. D",
    volume = "111",
    number = "5",
    pages = "056019",
    year = "2025"
}

@article{NANOGrav:2023gor,
    author = "Agazie, Gabriella and others",
    collaboration = "NANOGrav",
    title = "{The NANOGrav 15 yr Data Set: Evidence for a Gravitational-wave Background}",
    eprint = "2306.16213",
    archivePrefix = "arXiv",
    primaryClass = "astro-ph.HE",
    doi = "10.3847/2041-8213/acdac6",
    journal = "Astrophys. J. Lett.",
    volume = "951",
    number = "1",
    pages = "L8",
    year = "2023"
}

@article{NANOGrav:2023hde,
    author = "Agazie, Gabriella and others",
    collaboration = "NANOGrav",
    title = "{The NANOGrav 15 yr Data Set: Observations and Timing of 68 Millisecond Pulsars}",
    eprint = "2306.16217",
    archivePrefix = "arXiv",
    primaryClass = "astro-ph.HE",
    doi = "10.3847/2041-8213/acda9a",
    journal = "Astrophys. J. Lett.",
    volume = "951",
    number = "1",
    pages = "L9",
    year = "2023"
}

@article{EPTA:2023sfo,
    author = "Antoniadis, J. and others",
    collaboration = "EPTA",
    title = "{The second data release from the European Pulsar Timing Array - I. The dataset and timing analysis}",
    eprint = "2306.16224",
    archivePrefix = "arXiv",
    primaryClass = "astro-ph.HE",
    doi = "10.1051/0004-6361/202346841",
    journal = "Astron. Astrophys.",
    volume = "678",
    pages = "A48",
    year = "2023"
}

@article{EPTA:2023gyr,
    author = "Antoniadis, J. and others",
    collaboration = "EPTA, InPTA",
    title = "{The second data release from the European Pulsar Timing Array - V. Search for continuous gravitational wave signals}",
    eprint = "2306.16226",
    archivePrefix = "arXiv",
    primaryClass = "astro-ph.HE",
    doi = "10.1051/0004-6361/202348568",
    journal = "Astron. Astrophys.",
    volume = "690",
    pages = "A118",
    year = "2024"
}

@article{EPTA:2023xxk,
    author = "Antoniadis, J. and others",
    collaboration = "EPTA, InPTA",
    title = "{The second data release from the European Pulsar Timing Array - IV. Implications for massive black holes, dark matter, and the early Universe}",
    eprint = "2306.16227",
    archivePrefix = "arXiv",
    primaryClass = "astro-ph.CO",
    doi = "10.1051/0004-6361/202347433",
    journal = "Astron. Astrophys.",
    volume = "685",
    pages = "A94",
    year = "2024"
}

@article{EPTA:2023fyk,
    author = "Antoniadis, J. and others",
    collaboration = "EPTA, InPTA:",
    title = "{The second data release from the European Pulsar Timing Array - III. Search for gravitational wave signals}",
    eprint = "2306.16214",
    archivePrefix = "arXiv",
    primaryClass = "astro-ph.HE",
    doi = "10.1051/0004-6361/202346844",
    journal = "Astron. Astrophys.",
    volume = "678",
    pages = "A50",
    year = "2023"
}

@article{EPTA:2023akd,
    author = "Antoniadis, J. and others",
    collaboration = "EPTA, InPTA",
    title = "{The second data release from the European Pulsar Timing Array - II. Customised pulsar noise models for spatially correlated gravitational waves}",
    eprint = "2306.16225",
    archivePrefix = "arXiv",
    primaryClass = "astro-ph.HE",
    doi = "10.1051/0004-6361/202346842",
    journal = "Astron. Astrophys.",
    volume = "678",
    pages = "A49",
    year = "2023"
}

@article{Reardon:2023gzh,
    author = "Reardon, Daniel J. and others",
    title = "{Search for an Isotropic Gravitational-wave Background with the Parkes Pulsar Timing Array}",
    eprint = "2306.16215",
    archivePrefix = "arXiv",
    primaryClass = "astro-ph.HE",
    doi = "10.3847/2041-8213/acdd02",
    journal = "Astrophys. J. Lett.",
    volume = "951",
    number = "1",
    pages = "L6",
    year = "2023"
}

@article{Reardon:2023zen,
    author = "Reardon, Daniel J. and others",
    title = "{The Gravitational-wave Background Null Hypothesis: Characterizing Noise in Millisecond Pulsar Arrival Times with the Parkes Pulsar Timing Array}",
    eprint = "2306.16229",
    archivePrefix = "arXiv",
    primaryClass = "astro-ph.HE",
    doi = "10.3847/2041-8213/acdd03",
    journal = "Astrophys. J. Lett.",
    volume = "951",
    number = "1",
    pages = "L7",
    year = "2023"
}

@article{Zic:2023gta,
    author = "Zic, Andrew and others",
    title = "{The Parkes Pulsar Timing Array third data release}",
    eprint = "2306.16230",
    archivePrefix = "arXiv",
    primaryClass = "astro-ph.HE",
    doi = "10.1017/pasa.2023.36",
    journal = "Publ. Astron. Soc. Austral.",
    volume = "40",
    pages = "e049",
    year = "2023"
}

@article{Xu:2023wog,
    author = "Xu, Heng and others",
    title = "{Searching for the Nano-Hertz Stochastic Gravitational Wave Background with the Chinese Pulsar Timing Array Data Release I}",
    eprint = "2306.16216",
    archivePrefix = "arXiv",
    primaryClass = "astro-ph.HE",
    doi = "10.1088/1674-4527/acdfa5",
    journal = "Res. Astron. Astrophys.",
    volume = "23",
    number = "7",
    pages = "075024",
    year = "2023"
}

@article{LISACosmologyWorkingGroup:2022kbp,
    author = "Bartolo, Nicola and others",
    collaboration = "LISA Cosmology Working Group",
    title = "{Probing anisotropies of the Stochastic Gravitational Wave Background with LISA}",
    eprint = "2201.08782",
    archivePrefix = "arXiv",
    primaryClass = "astro-ph.CO",
    doi = "10.1088/1475-7516/2022/11/009",
    journal = "JCAP",
    volume = "11",
    pages = "009",
    year = "2022"
}

@article{LISACosmologyWorkingGroup:2019mwx,
    author = "Belgacem, Enis and others",
    collaboration = "LISA Cosmology Working Group",
    title = "{Testing modified gravity at cosmological distances with LISA standard sirens}",
    eprint = "1906.01593",
    archivePrefix = "arXiv",
    primaryClass = "astro-ph.CO",
    reportNumber = "LISA CosWG-19-01; IFT-UAM-CSIC-19-79, LISA CosWG-19-01",
    doi = "10.1088/1475-7516/2019/07/024",
    journal = "JCAP",
    volume = "07",
    pages = "024",
    year = "2019"
}

@article{LISA:2022kgy,
    author = "Arun, K. G. and others",
    collaboration = "LISA",
    title = "{New horizons for fundamental physics with LISA}",
    eprint = "2205.01597",
    archivePrefix = "arXiv",
    primaryClass = "gr-qc",
    doi = "10.1007/s41114-022-00036-9",
    journal = "Living Rev. Rel.",
    volume = "25",
    number = "1",
    pages = "4",
    year = "2022"
}

@article{LISA:2017pwj,
    author = "Amaro-Seoane, Pau and others",
    collaboration = "LISA",
    title = "{Laser Interferometer Space Antenna}",
    journal = {arXiv e-prints},
    eprint = "1702.00786",
    archivePrefix = "arXiv",
    primaryClass = "astro-ph.IM",
    month = "2",
    year = "2017"
}

@article{LIGOScientific:2016aoc,
    author = "Abbott, B. P. and others",
    collaboration = "LIGO Scientific, Virgo",
    title = "{Observation of Gravitational Waves from a Binary Black Hole Merger}",
    eprint = "1602.03837",
    archivePrefix = "arXiv",
    primaryClass = "gr-qc",
    reportNumber = "LIGO-P150914",
    doi = "10.1103/PhysRevLett.116.061102",
    journal = "Phys. Rev. Lett.",
    volume = "116",
    number = "6",
    pages = "061102",
    year = "2016"
}

@article{Punturo:2010zz,
    author = "Punturo, M. and others",
    editor = "Ricci, Fulvio",
    title = "{The Einstein Telescope: A third-generation gravitational wave observatory}",
    doi = "10.1088/0264-9381/27/19/194002",
    journal = "Class. Quant. Grav.",
    volume = "27",
    pages = "194002",
    year = "2010"
}

@article{Sathyaprakash:2012jk,
    author = "Sathyaprakash, B. and others",
    editor = "Hannam, Mark and Sutton, Patrick and Hild, Stefan and van den Broeck, Chris",
    title = "{Scientific Objectives of Einstein Telescope}",
    eprint = "1206.0331",
    archivePrefix = "arXiv",
    primaryClass = "gr-qc",
    doi = "10.1088/0264-9381/29/12/124013",
    journal = "Class. Quant. Grav.",
    volume = "29",
    pages = "124013",
    year = "2012",
    note = "[Erratum: Class.Quant.Grav. 30, 079501 (2013)]"
}

@article{Sathyaprakash:2009xt,
    author = "Sathyaprakash, B. S. and Schutz, B. F. and Van Den Broeck, C.",
    title = "{Cosmography with the Einstein Telescope}",
    eprint = "0906.4151",
    archivePrefix = "arXiv",
    primaryClass = "astro-ph.CO",
    doi = "10.1088/0264-9381/27/21/215006",
    journal = "Class. Quant. Grav.",
    volume = "27",
    pages = "215006",
    year = "2010"
}

@article{Reitze:2019iox,
    author = "Reitze, David and others",
    title = "{Cosmic Explorer: The U.S. Contribution to Gravitational-Wave Astronomy beyond LIGO}",
    eprint = "1907.04833",
    archivePrefix = "arXiv",
    primaryClass = "astro-ph.IM",
    reportNumber = "LIGO-P1900316",
    journal = "Bull. Am. Astron. Soc.",
    volume = "51",
    number = "7",
    pages = "035",
    year = "2019"
}

@article{Mpetha:2022xqo,
    author = "Mpetha, Charlie T. and Congedo, Giuseppe and Taylor, Andy",
    title = "{Future prospects on testing extensions to \ensuremath{\Lambda}CDM through the weak lensing of gravitational waves}",
    eprint = "2208.05959",
    archivePrefix = "arXiv",
    primaryClass = "astro-ph.CO",
    doi = "10.1103/PhysRevD.107.103518",
    journal = "Phys. Rev. D",
    volume = "107",
    number = "10",
    pages = "103518",
    year = "2023"
}

@article{Evans:2021gyd,
    author = "Evans, Matthew and others",
    title = "{A Horizon Study for Cosmic Explorer: Science, Observatories, and Community}",
    journal = {arXiv e-prints},
    eprint = "2109.09882",
    archivePrefix = "arXiv",
    primaryClass = "astro-ph.IM",
    reportNumber = "CE-P2100003-v7, Cosmic Explorer technical report CE-P2100003-v6",
    month = "9",
    year = "2021"
}

@book{Ortin:2015hya,
    author = "Ortin, Tomas",
    title = "{Gravity and Strings}",
    edition = "2nd ed.",
    doi = "10.1017/CBO9781139019750",
    isbn = "978-0-521-76813-9, 978-0-521-76813-9, 978-1-316-23579-9",
    publisher = "Cambridge University Press",
    series = "Cambridge Monographs on Mathematical Physics",
    month = "7",
    year = "2015"
}

@ARTICLE{Virgo,
       author = {T. {Accadia} and others},
        title = "{Virgo: a laser interferometer to detect gravitational waves}",
      journal = {Journal of Instrumentation},
         year = 2012,
        month = mar,
       volume = {7},
       number = {3},
        pages = {3012},
          doi = {10.1088/1748-0221/7/03/P03012},
       adsurl = {https://ui.adsabs.harvard.edu/abs/2012JInst...7.3012A}
}

@ARTICLE{Kagra,
       author = {{Kagra Collaboration}},
        title = "{KAGRA: 2.5 generation interferometric gravitational wave detector}",
      journal = {Nature Astronomy},
         year = 2019,
        month = jan,
       volume = {3},
        pages = {35-40},
          doi = {10.1038/s41550-018-0658-y},
archivePrefix = {arXiv},
       eprint = {1811.08079},
 primaryClass = {gr-qc},
       adsurl = {https://ui.adsabs.harvard.edu/abs/2019NatAs...3...35K}
}

@ARTICLE{LISA,
       author = {{Colpi}, Monica and others},
        title = "{LISA Definition Study Report}",
      journal = {arXiv e-prints},
         year = 2024,
        month = feb,
          eid = {arXiv:2402.07571},
        pages = {arXiv:2402.07571},
          doi = {10.48550/arXiv.2402.07571},
archivePrefix = {arXiv},
       eprint = {2402.07571},
 primaryClass = {astro-ph.CO},
       adsurl = {https://ui.adsabs.harvard.edu/abs/2024arXiv240207571C}
}

@article{TianQin,
    author = "Mei, Jianwei and others",
    collaboration = "TianQin",
    title = "{The TianQin project: current progress on science and technology}",
    eprint = "2008.10332",
    archivePrefix = "arXiv",
    primaryClass = "gr-qc",
    doi = "10.1093/ptep/ptaa114",
    journal = "PTEP",
    volume = "2021",
    number = "5",
    pages = "05A107",
    year = "2021"
}

@article{PhysRev.131.435,
  title = {Gravitational Radiation from Point Masses in a Keplerian Orbit},
  author = {Peters, P. C. and Mathews, J.},
  journal = {Phys. Rev.},
  volume = {131},
  issue = {1},
  pages = {435--440},
  numpages = {0},
  year = {1963},
  month = {Jul},
  publisher = {American Physical Society},
  doi = {10.1103/PhysRev.131.435},
  url = {https://link.aps.org/doi/10.1103/PhysRev.131.435}
}

@article{LISAConsortiumWaveformWorkingGroup:2023arg,
    author = "Afshordi, Niayesh and others",
    collaboration = "LISA Consortium Waveform Working Group",
    title = "{Waveform modelling for the Laser Interferometer Space Antenna}",
    eprint = "2311.01300",
    archivePrefix = "arXiv",
    primaryClass = "gr-qc",
    doi = "10.1007/s41114-025-00056-1",
    journal = "Living Rev. Rel.",
    volume = "28",
    number = "1",
    pages = "9",
    year = "2025"
}

@article{Branchesi:2023mws,
    author = "Branchesi, Marica and others",
    title = "{Science with the Einstein Telescope: a comparison of different designs}",
    eprint = "2303.15923",
    archivePrefix = "arXiv",
    primaryClass = "gr-qc",
    reportNumber = "ET-0084A-23",
    doi = "10.1088/1475-7516/2023/07/068",
    journal = "JCAP",
    volume = "07",
    pages = "068",
    year = "2023"
}

\end{document}